\documentclass[oldversion]{lfiaa}
\usepackage{epsfig}
\usepackage{subfigure}
\usepackage{amsmath,amssymb}
\usepackage{graphicx}
\usepackage{natbib}
\begin{document}
   \title{Planck-LFI: Design and Performance of the 4 Kelvin Reference Load Unit}

   \author{
L. Valenziano\inst{1} 
	\and 
F. Cuttaia\inst{1}
	\and
A. De Rosa\inst{1}
	\and
L. Terenzi\inst{1}
	\and
A. Brighenti\inst{1}
	\and
G.P. Cazzola\inst{1}
	\and
A. Garbesi\inst{2}
	\and
S. Mariotti\inst{3}
	\and
G. Orsi\inst{1}
	\and
L. Pagan\inst{5}
	\and
F. Cavaliere\inst{6}
	\and
M. Biggi\inst{4}
	\and
R. Lapini\inst{4}
	\and
E. Panagin\inst{4}
	\and
P. Battaglia\inst{4}
	\and
R.C. Butler\inst{1}
	\and
M. Bersanelli\inst{6}
	\and
O. D'Arcangelo\inst{7}
	\and
S. Levin\inst{8}
	\and
N. Mandolesi\inst{1}
	\and
A. Mennella\inst{6}
	\and
G. Morgante\inst{1}
	\and
G. Morigi\inst{1}
	\and
M. Sandri\inst{1}
	\and
A. Simonetto\inst{7}
	\and
M. Tomasi\inst{9}
	\and 
F. Villa\inst{1}
	\and
M. Frailis\inst{10}
	\and
S. Galeotta\inst{11}
	\and
A. Gregorio\inst{10,11}
	\and
R. Leonardi\inst{12}
	\and
S.R. Lowe\inst{13}
	\and
M. Maris\inst{10}
	\and
P. Meinhold\inst{12}
\and
L. Mendes\inst{14}
\and
L. Stringhetti\inst{1}
	\and 
A. Zonca\inst{9}
	\and
A. Zacchei\inst{10}
}

   \offprints{L. Valenziano \email{valenziano@iasfbo.inaf.it}}

   \institute{   
Istituto di Astrofisica Spaziale e Fisica Cosmica - Bologna, INAF, via P. Gobetti, 101 -- I40129 Bologna, Italy
	\and
Istituto per la Sintesi Organica e la Fotoreattivit\`a, CNR, via P. Gobetti, 101 -- I40129 Bologna, Italy
	\and
Istituto di Radioastronomia, INAF, via P. Gobetti, 101 -- I40129 Bologna, Italy
	\and
Officine Pasquali, via Palazzo de' Diavoli 124B, -- I50142 Firenze, Italy
	\and
Thales Alenia Space Italia, Sede di Milano, S.S. Padana Superiore, 290 -- I20090 Vimodrone, Italy
	\and
Universit\`a degli Studi di Milano, Via Celoria 16, -- I20133 Milano, Italy
   \and
Istituto di Fisica del Plasma, CNR, Via Roberto Cozzi, 53 -- I20125 Milano, Italy
	\and
Jet Propulsion Laboratory, 4800 Oak Grove Drive, CA91109 -- Pasadena, USA
	\and
Istituto di Astrofisica Spaziale e Fisica Cosmica - Milano, INAF, via E. Bassini, 15 -- I20133 Milano, Italy
	\and
Osservatorio Astronomico di Treste, INAF, via G.B. Tiepolo, 11 -- I34143 Trieste, Italy
	\and
Universit\'a degli studi di Trieste, Dipartimento di Fisica, via A. Valerio, 2 -- I34127 Trieste, Italy
	\and
Department of Physics, University of California, Santa Barbara 2225 Broida Hall -- CA 93106 Santa Barbara, USA 
	\and
Jodrell Bank Centre for Astrophysics, Alan Turing Building, The University of Manchester, Manchester, M13 9PL, UK
	\and
ESA - European Space Agency, Keplerlaan 1, NL 2200 AG Noordwijk, The Netherlands
}


   \date{Received June 23,2009; accepted August 5, 2009}
 
  \abstract
 { 
The LFI radiometers use a pseudo-correlation design where the signal from the sky is continuously compared with a stable reference signal, provided by a cryogenic reference load system. The reference unit is composed by small pyramidal horns, one for each radiometer, 22 in total, facing small absorbing targets, made of a commercial resin ECCOSORB CR\texttrademark, cooled to $\sim$4.5 K. Horns and targets are separated by a small gap to allow thermal decoupling. Target and horn design is optimized for each of the LFI bands, centered at 70, 44 and 30 GHz. Pyramidal horns are either machined inside the radiometer 20K module or connected via external electro-formed bended waveguides. The requirement of high stability of the reference signal imposed a careful design for the radiometric and thermal properties of the loads. Materials used for the manufacturing have been characterized for thermal, RF and mechanical properties. We describe in this paper the design and the performance of the reference system. 
}


   \maketitle

\small\begin{quote}\begin{center}\rule{240pt}{1pt}\\\vspace{6pt} {\bf Remark to the A{\sc r}X{\sc i}V version}\end{center}\noindent
This is an author-created, un-copyedited version of an article published in JINST.IOP Publishing Ltd is not responsible for any errors or omissions in this version of the manuscript or any version derived from it.The present version is derived from the latest version of the paperbefore final acceptance from JINST, thus itcould have some minor differences in phrasing, spelling and style with respect to the published version.The definitive publisher authenticated version is available online at:\end{quote}
\begin{center}
http://www.iop.org/EJ/journal/-page=extra.proc5/jinst \\
Reference : 2009 JINST 4 T12006  \\
DOI: 10.1088/1748-0221/4/12/T12006 
\vspace{6pt}\rule{240pt}{1pt}
\end{center}
%

\section{Introduction}

Planck \citep{2001IAUS..204..493T,2002AdSpR..30.2123M,2004cosp...35.4559P,2005IAUS..201...86T,2006msu..conf...35T} is the third generation mission devoted to produce the ultimate image the Cosmic Microwave Background (CMB) anisotropies. 
Planck will be placed on a Lissajou orbit around the second Lagrangian point of the Earth-Sun system. By following our planet in its revolution motion, it will always be oriented in the anti-Sun direction, spinning around its z-axis at 1 r.p.m. The telescope, whose line-of-sight is at $\sim$ 90 deg from the spin axis, will allow all detectors in the focal plane to scan almost the whole sky in six months.

Two instruments share the focal surface of a 1.5 mirror off-axis telescope \citep{2000A&AS..145..323M, Tauber09b}: the High Frequency Instrument (HFI) \citep{2003NewAR..47.1017L,2003SPIE.4850..730L,Lamarre09} and the Low Frequency Instrument (LFI) \citep{2002AIPC..609..144V,2002AIPC..616..193M,2004AIPC..703..401M,2004MSAIS...5..411S,2007NewAR..51..287V,Mandolesi2009, Bersanelli2009}. Planck was successfully launched on May, 14, 2009 on an Ariane 5 rocket from Kourou spaceport in French Guiana.

The LFI is an array of 22 pseudo-correlation radiometers \citep{Bersanelli2009}, based on InP HEMT Low Noise  Amplifiers (LNAs). Each radiometer is continuously comparing the signal from the sky with a stable reference signal provided by a unit, called 4K Reference Load (4KRL). It is composed of an assembly of small absorbing loads, thermally and mechanically connected to the HFI shield at a temperature of $\sim$4.5 K. Each load is composed by a termination-like target, made of \footnote{ECCOSORB is a TradeMark of Emerson and Cuming} ECCOSORB\texttrademark, facing a small pyramidal horn, connected to the 20K radiometer reference arm. 

The Planck instruments are cooled by a complex active system: a Hydrogen Sorption Cooler \citep{Bhandari04:planck_sorption_cooler, Morgante2009} keeps the LFI at $\sim$ 20K and acts as pre-cooling stage for the HFI cryo-chain. This latter is composed by a mechanical 4K cooler \citep{Bradshaw}, which brings the HFI outer shield, where the 4KRL unit is mounted, at approximately 4.5 K. Subsequent stages are composed a diluition cooler (internally pre-cooled at 1.6K), which brings the HFI bolometers to T$\sim$ 100 mK \citep{Triqueneaux}.

An overview of the 4KRL unit is presented in Section \ref{section:overview}; in Section \ref{section:requirements}  we describe how the requirements for the 4KRL were derived; in Section \ref{section:design} the mechanical, radiometric and thermal design of the 4KRL unit is reported; we describe in Section \ref{section:performance} the measured performance of the unit and we compare them with the requirements. Finally, some material properties are reported in the Appendix. 

\section{An overview of the 4K Reference Load Unit \label{section:overview}}

\begin{figure}
	\centering
		\includegraphics[width=\linewidth]{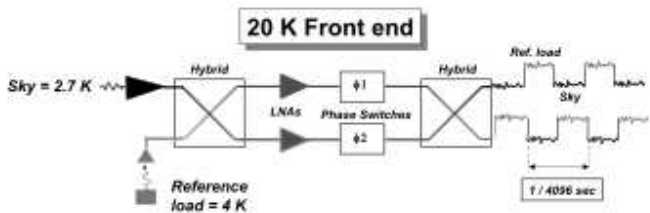}
	\caption{Schematic view of a LFI Front-End Module (adapted from \cite{Mennella09a}).}
	\label{fig:fem_sketch}
\end{figure}

LFI uses a pseudo-correlation receiver design \citep{Bersanelli2009}. This radiometer concept is chosen to maximize the stability of the instrument by reducing the effect of non-white noise generated in the radiometer itself.
In this scheme (see Figure \ref{fig:fem_sketch}), the difference between the inputs to each chains (the signal from the telescope and the signal from the 4KRL) is continuously being observed. To minimize the 1/f noise of the radiometers, the reference blackbody temperature should be as close as possible to the sky temperature ($\sim$3K). To remove the effect of instability in the back-end amplifiers and detectors diodes, it is necessary to modulate the signals using phase switches. This design was chosen over a much simpler total-power scheme (consisting of one of the two parallel chains) because the latter exhibits inadequate gain stability at time scales larger than a few seconds (to be compared with the spin period, 60s). 

The purpose of the 4KRL is to provide the radiometers with a low input offset (the radiometric temperature difference between the sky and the reference load). Reducing the input offset reduces the minimum achievable radiometer knee frequency for a given amplifier fluctuation spectrum \citep{2002A&A...391.1185S}.  This minimum achievable knee frequency assumes perfect phase and gain matching in the two "legs" of the radiometer and assumes other ideal characteristics in radiometer components. An ideal reference load temperature would match the sky temperature (approximately 2.7 K), but there is no convenient spacecraft source of 2.7 K with sufficient cooling power. A gain modulation factor is introduced to compensate for this effect \citep{2003A&A...410.1089M}. Moreover, minimising the input offset reduces the potential of multiplicative systematic effects to contaminate the measurement \citep{2002A&A...391.1185S}. 

\begin{figure}
	\centering
		\includegraphics[width=\linewidth]{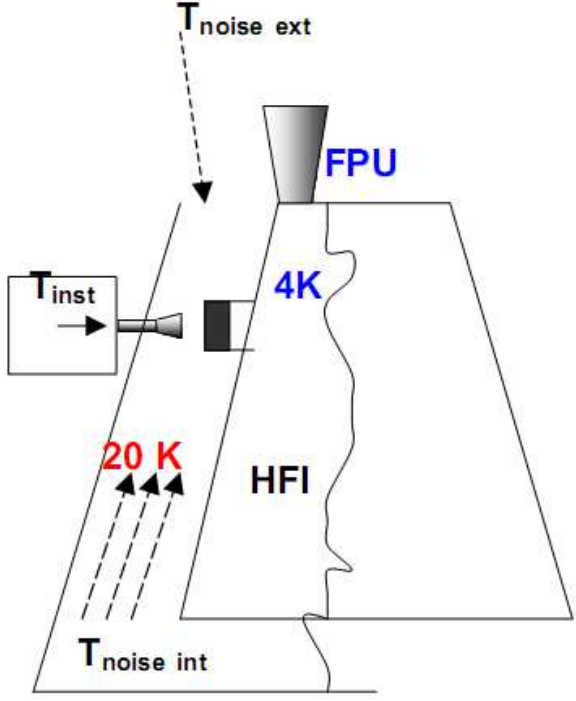}

	\caption{Schematic representation signals reaching the 4KRL. One  feed placed in the FPU looking at the sky, the 4K HFI shield, one 4KRL  (connected to the 4K shield) and one 4K reference horn (connected to the LFI main  frame at 20K) are shown. On the left, a box representing the RF chain behind the horn is displayed.}
	\label{fig:spo}
\end{figure}

One of the main requirements for the 4KRL design was to minimize the heat load on the HFI to a value lower than 1 mW. Safety considerations (a thermal short between the two instruments will prevent the HFI to work) lead to mechanically decouple the loads, mounted on the HFI external shield, from the LFI radiometers, at 20K. A system composed by an absorbing target (Reference Target - RT) at 4K facing a receiving horn (Reference Horn - RH), connected to the radiometer at 20K, was selected (Figure \ref{fig:spo}).

\section{4KRL requirements\label{section:requirements}}

The design of the 4KRL unit is derived from performance required to the Low Frequency Instrument to obtain the Planck scientific objectives \citep{Mandolesi2009, Bersanelli2009}. We made a budget in which the overall requirements are split in the various contributions. We then set limits to each component to comply with the total requirement (both temperature stability and absolute temperature). The main quantities related to the 4KRL performance are the $1/f$ knee frequency, which is linked to difference between the reference and sky temperature \citep{2002A&A...391.1185S}, the temperature stability \citep{2002AIPC..616..229M} and the 4KRL sensitivity to spurious RF components that could contaminate the reference signal.
The requirements apply to the signal as measured at the input of the hybrid coupler in the FEM, where it is "mixed" with the signal from the sky. The design then has to be compliant with the allocated volume and mass for the unit and with thermo-mechanical constraints due to differential contraction when the reference load is cooled down to 4K.
Moreover the design must be robust in term of safety, especially as far as vibrations at launch are concerned. The framework in which the 4KRL was developed is based on all these considerations.

The 4KRL RF design is intimately linked to its thermal design: the requirement on the maximum heat load on the HFI lead to a horn-target design, therefore allowing external signals to enter the radiometer reference arm (we call this effect {\it leakage}). This solution implies the presence of a gap in the radiometer reference arm, through which external spurious signals can leak in the radiometers.
We must also consider the non-idealities of the reference system: mismatching between the load and the reference horn, losses in the reference arm.
The need to dump the reference signal fluctuations imposed a trade-off on the thermal coupling between the reference targets and the heat sink, the HFI shields, impacting on the minimum temperature achivable by the loads.

We report here some useful definitions which are used in specifying the requirements. We express here the power in antenna temperature, $T_A$, in the Rayleigh Jeans approximation, and measured in Kelvin. 

Ideally, the power that the radiometer receives (at the hybrid  coupler) from a reference target at physical temperature $T_{RT}$ can be expressed as
\begin{equation}
T_{Load} = \epsilon_{RT} T_{RT}
\end{equation}

where $\epsilon_{RT} $ is the emissivity of the reference target, and can be expressed, in the case of an ideal system, in terms of load reflectivity as $ \epsilon_{RT} = (1 - R_{RT})$. 

\begin{figure}
	\centering
		\includegraphics[width=\linewidth]{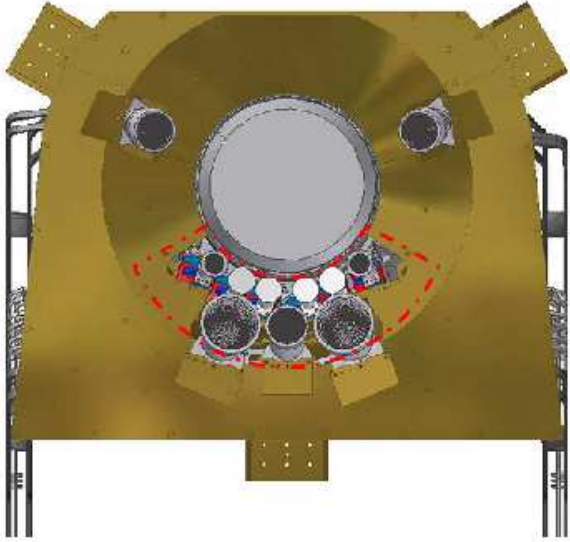}
	\label{fig:mainframe}
	\caption{Main frame top view: the red dot-dashed line highlights the gap around the feeds, allowing radiation from outside to enter the cavity}
\end{figure}

We identified the following terms contributing to the overall effective temperature: 

\begin{equation}
T_{4KRL} = T_{Load} + T_{leak} + T_{refl} + T_{loss}
\label{eq:t4k}
\end{equation}

where
\begin{eqnarray}
&&T_{Load} = T_{RT} \cdot (1 - R_{RT})  \\
&&T_{leak} = [T_{int}  \cdot \epsilon_{int} + \nonumber\\
&& + (T^{sky}_{ext} + T^{payload}\cdot \epsilon_{ext}) \cdot SPO_{ext}] \cdot SPO_{int} \nonumber\\
&&T_{refl} = T^{rad}_N \cdot R_{RT} \nonumber\\
&&T_{loss} = T_{RH} \cdot L_{4KRL} \nonumber\\
\label{eq:t4k_arr}
\end{eqnarray}

\begin{itemize}

\item $T_{RT}$ is the effective antenna temperature of the blackbody load, taking into account: HFI 4K shield temperature at the interface of the reference loads and the thermal properties of the load and of the thermal link.

\item $(1 - R_{RT})$ is the load emissivity, which include also the horn contribution.
\item $T_{leak}$ is radiation leaking into the reference horn from the horn-target gap. 
\item $T_{int}$ is the temperature of the LFI-HFI cavity.
\item $SPO_{int,ext}$ is the {\it spillover} dumping factor of external signals entering in the cavity (ext) and in the horn-target gap (int).
\item $T_{refl}$ is the amplifier noise temperature, radiated in the radiometer reference arm and reflected back from targets.
\item $T^{rad}_N$ is the radiometer noise temperature.
\item $T_{loss}$ is the contribution due to the ohmic loss of the reference horn (and the external reference waveguide for the 30 and 44 GHz).
\item $R_{RT}$ is the mismatch at the horn aperture, when it is facing the target.
\item $\epsilon_{int,ext}$ is the emissivity of the {\it internal} and the {\it external} environment, where the former refers to the cavity between the LFI and the HFI and the latter to the payload environment, inside the radiation shield surrounding the telescope.
\item $WG$: refers to the waveguide connecting each Reference Horn ($RH$) to the radiometer reference arm.
\end{itemize}

It is evident that the load {\it effective} temperature, $T_{4KRL}$ depends on target physical temperature and on the radiative behavior of the horn-target ensemble. We need the dominant contribution to be $T_{RT}$ and we need to set the requirements to design a unit which makes the other contribution negligible and well understood.

The critical areas are:
\begin{itemize}
	\item The mismatching RL-RH, $R_{RT}$: it causes power loss, and take the load away from having a perfect blackbody spectrum. It is strictly related to the reference horn geometry, load shape and material.
	\item The Spillover $SPO$: it causes a power loss from the load and power entrance from external (coming from instrument or sky). It depends on the RH near field pattern, on the intrinsic directional RL emissivity, on the RH-RL gap thickness.
	\item The power loss due to ohmic effects in the RH+WG unit. They attenuate the reference signal  by $L_{4KRL} \cdot T_{RT}$  and increase the antenna temperature by a larger quantity $L_{4KRL} \cdot T_{RH}$.
	  	
\end{itemize}

While the load absolute {\it effective} temperature is relevant for the $1/f$ knee frequency, reference load signal fluctuations are by far more important, since they could mimic a sky signal. 

Requirements on fluctuations are stringent: the 4KRL unit is designed to have spin synchronous signal (SS) fluctuations less than 1 $\mu K$ per 30 arcmin sq. pixel on the final maps, after consolidated software removal techniques \citep{Bersanelli2009}.

Changes in the physical temperature of the reference load show up directly as errors in the output signal from the radiometer. The 4KRL signal stability requirement depends on the time scale of the variations. Variations on time scales long compared to the 60 s spin period are suppressed when the multiple observations of a given spot on the sky are combined; variations on time scales short compared to the radiometer sampling time are averaged out in a single sample. It is therefore necessary to distinguish three different cases: random, spin synchronous and sorption cooler synchronous temperature fluctuations. Each of the different contributors to temperature fluctuations will be analyzed considering these three cases. Random fluctuations are uncorrelated and, in general, add quadratically. SS  fluctuations, in the worst case, add linearly.

If we differentiate Eq. \ref{eq:t4k}, we can evaluate the different contributions:

\begin{equation}
\Delta T_{4KRL} = \Delta T_{Load} + \Delta T_{leak} + \Delta T_{refl} + \Delta T_{loss}
\label{eq:deltat4k}
\end{equation}

We assume the third term neglible, since the radiometer $T_N$ does not significantly fluctuate with time and we neglect also the last one, since it is balanced, to the first order, by the fluctuation on the radiometer sky arm. $\Delta T_{Load}$ is directly related to the HFI shield fluctuations, mainly induced by the Sorption Cooler  \citep{ Bhandari04:planck_sorption_cooler, Morgante2009} and the 4K Cooler temperature oscillations \citep{Lamarre09}. These can be damped by reducing the thermal link between the targets and the heat sink to an extent limited by mechanical contraints and by the maximum temperature allowed for the loads (i.e. the larger is the damping, the higher the load temperature). 

Requirements on temperature oscillations of the HFI 4K shield are different for different frequency intervals: the most stringent at low frequency (close to spin frequency). We therefore set complementary requirements on dumping factors between HFI interface to the 4KRL and the front surface of the load

\begin{equation}
\Delta T_{RT} = D_f \cdot \Delta T_{HFI}
\end{equation}

The required values for $D_f$ are $<$0.1 and $<$ 0.9 at a period of 60s and between 600s and 1000s, respectively.

The term $ \Delta T_{leak} $ is related to fluctuating signals entering the horn-target gap. They can be separated in two contributions: fluctuations of the cavity between the LFI and the HFI, where loads are located; fluctuations coming from telescope area and from the sky. These must be damped by assuming worst case values for $SPO_{int}$ and $SPO_{ext}$, respectively \citep{Bersanelli2009}.

Starting from top-level requirements, we derived requirements on each term contributing to equation \ref{eq:deltat4k}.

The most critical SS signal at the LFI frequency is the CMB dipole, of the order of few mK \citep{1996ApJ...473..576F}: it must be dumped at a level below $\sim \mu$K. This requires the total SPO factor to be $\le$ -40 dB. Since SS signals in the LFI-HFI cavity are expected, in the worst case assumption, lower than few $\mu$K, we set:

\begin{itemize}
	\item $SPO_{int} \leq -20dB$
	\item $SPO_{ext} \leq -20dB$
\end{itemize}

$T_{RT}$ should be known with an accuracy comparable to the knowledge of the CMB absolute temperature \citep{1996ApJ...473..576F}. Therefore we set:

\begin{itemize}
	\item $R_{RT} \leq -20dB$
\end{itemize}

In order to balance the two radiometer arms, the Insertion Loss $L_{4KRL}$ in the reference arm needs to be of the same order of magnitude of the sky arm one. We then require:

\begin{itemize}
	\item $L_{4KRL} \le 0.15 dB$
\end{itemize}

The requirement on the absolute temperature is $T_{4KRL} < 5K$, so to have a $1/f$ knee frequency lower than $f_{knee} < 0.016 Hz$, corresponding to a spin period of 60 seconds.
Moreover, the load need to operate over the whole radiometer bandwidth, 20\% around each of the LFI frequencies (30, 44 and 70 GHz).

\section{Design and manufacturing of the 4KRL unit\label{section:design}}

While the first design constraint is the maximum thermal load the 4KRL may dissipate on the HFI,
another relevant one is the location and the orientation of the LFI FEMs. They are placed along two almost concentric circles in the focal plane, each sky horn with a different orientation along its axis for polarization (see \cite{Sandri2009, Leahy2009, Bersanelli2009}). The 4KRL must also allow the integration of the HFI inside the LFI with all the loads already mounted. 

\begin{figure}
	\centering
		\includegraphics[width=\linewidth]{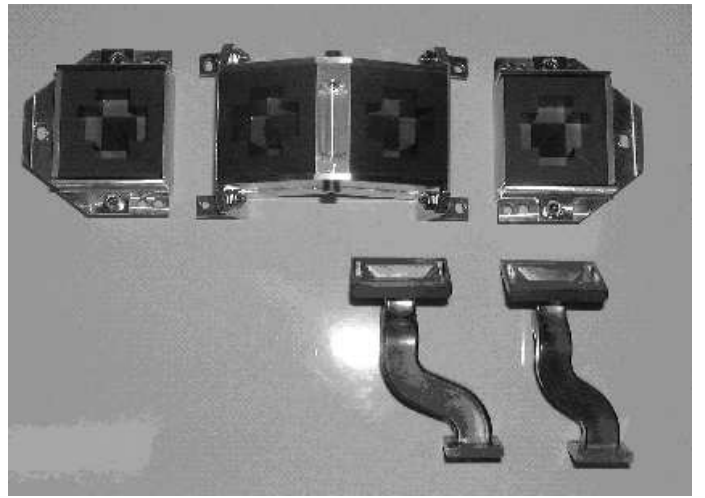}
	\caption{30GHz load. See Tables \ref{table:rt_data} and \ref{table:rh_data} for dimensions.}
	\label{fig:30GHz}
\end{figure}

\begin{figure}
	\centering
		\includegraphics[width=\linewidth]{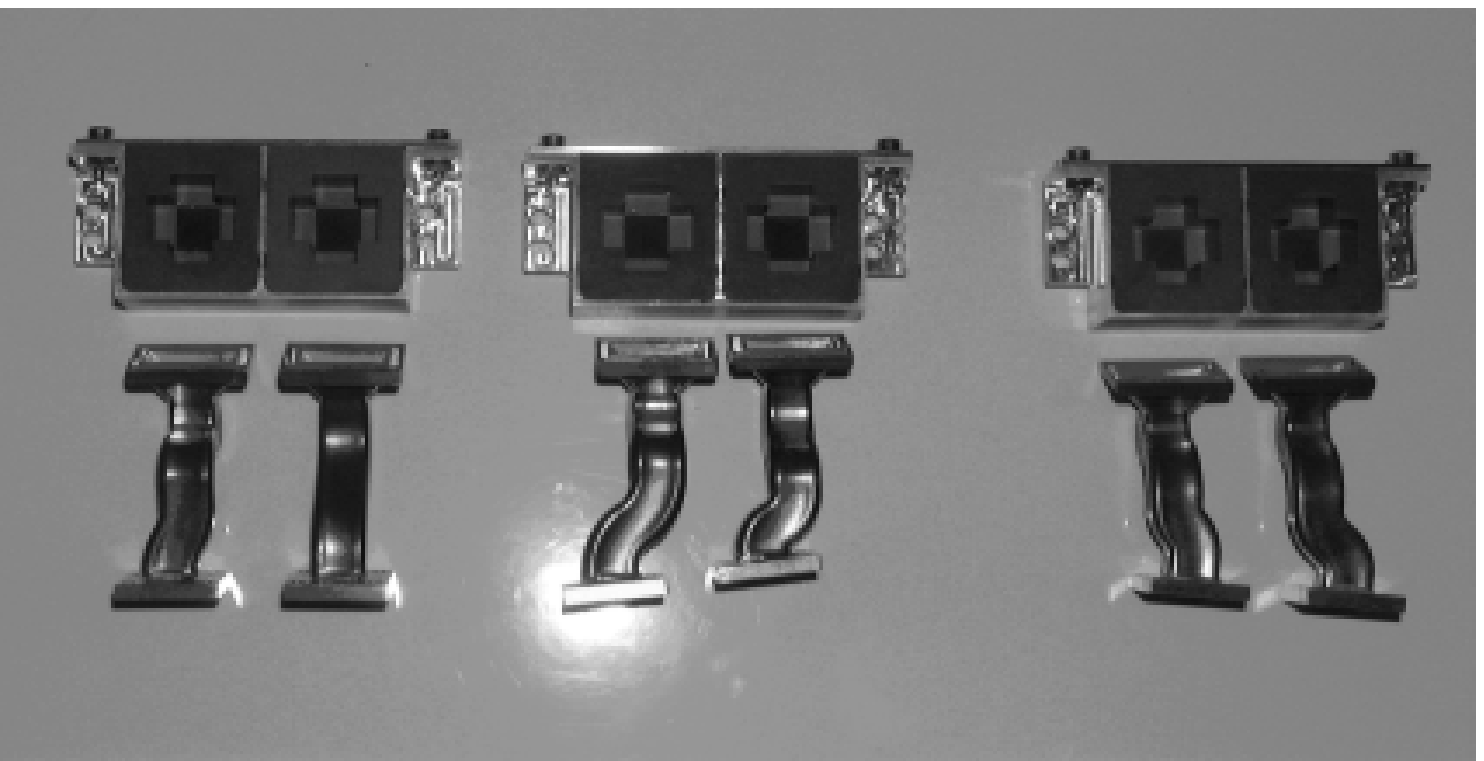}
	\caption{44GHz load. See Tables \ref{table:rt_data} and \ref{table:rh_data} for dimensions.}
	\label{fig:44GHz}
\end{figure}

\begin{figure}
	\centering
		\includegraphics[width=\linewidth]{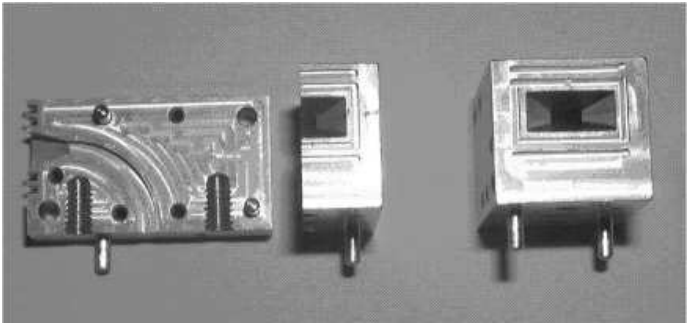}
	\caption{70 GHz reference horn used for 4KRL testing. The waveguide connecting to the hybryd and grooves around the aperture are visible. Real horns are internal to FEMs \citep{Varis2009}. See Tables \ref{table:rt_data} and \ref{table:rh_data} for dimensions.}
	\label{fig:70GHz}
\end{figure}

Reference horns need to be small compared to the wavelength: this ruled out conical, corrugated horns, also because this design would have needed a transition from circular to rectangular waveguide, to match the apertures of hybrid. Target also need to be small and placed in the very near field of the reference horns to reduce the leak from the gap. The requirement to operate over the full radiometer bandwidth ruled out a resonant load, intrinsically narrow-band. The conceptual design  is therefore based on small absorbing targets, mounted inside a metal enclosure ("case") to confine the radiation (see Figure \ref{fig:30GHz}, \ref{fig:44GHz}). Their shape is optimized to reduce the reflectivity. 
Cases, supported by an Al structure, are mounted on the HFI using Stainless Steel thermal decouplers ("washers"), which allows to carefully control the thermal link behavior. 
The reference pyramidal horns are derived from waveguide flares and optimized coupled to the targets (see the 70 GHz reference horn in Figure \ref{fig:70GHz}). The spacing between the horn and the targets is set to a nominal value of 1.5 mm. Particular care is posed in minimizing the signal {\it leaking} through the horn-target gap. 


\begin{figure}
	\centering
		\includegraphics[width=\linewidth]{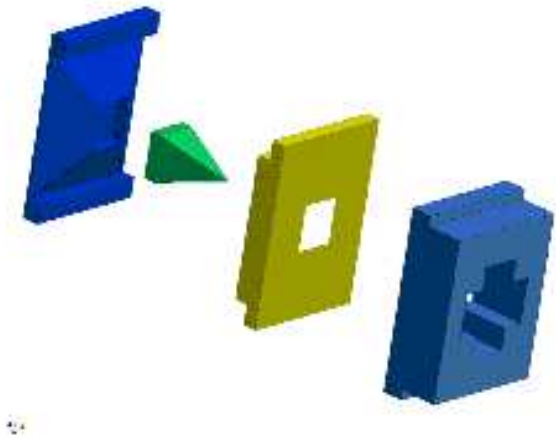}
		
	\caption{Exploded view of a 70 GHz target. The back part ({\it base}) is separated, in the 70GHz targets, in two sub-parts (blue and yellow). The pyramid (green) and the front part ({\it cross}, light blue) are visible.}
	\label{fig:target}
\end{figure}

\subsection{RF design}

\begin{figure}[htb]
	\centering
		\includegraphics{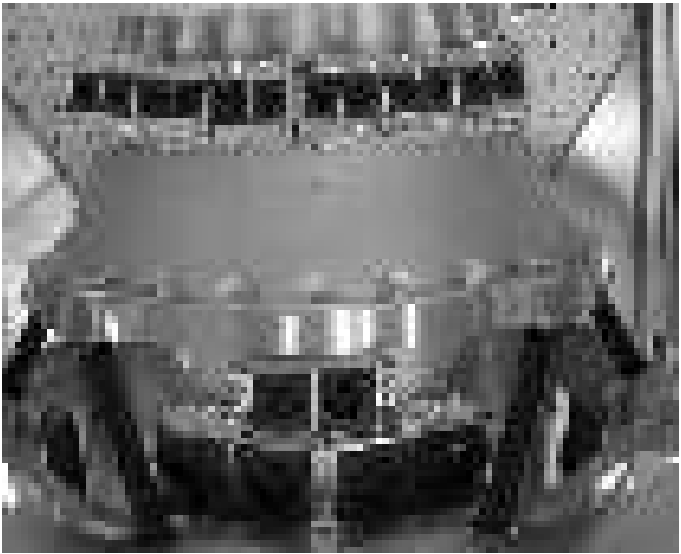}
	\caption{The 4K Reference Load unit on the HFI. The upper sector is mounted on the HFI cone. Note the reference target orientation, different for each 'twin' load. The 30GHz targets are shown in the lower sector. Reference horns are centered on each target, once the HFI is mated with the LFI. One of the 44 GHz target is visible, while the other two are on the back side of the HFI 4K shield, separated by 120$^{\circ}$.}
	\label{fig:4K_montato}
\end{figure}

The LFI is composed by 11 feed-horns, each one feeding two radiometers, one for each polarization \citep{Bersanelli2009}. The 4KRL unit is therefore composed by 22 reference loads (each one formed by a RT and a RH: 12, 6 and 4 for the 70, 44 and 30 GHz LFI radiometers, respectively), assembled in 'twin' sub-units, shown in Figures \ref{fig:30GHz},\ref{fig:44GHz} and \ref{fig:4K_montato} (70 GHz), one for each of the LFI Front-End Modules. 

The development of a detailed RF design followed an iterative process: horns and targets were designed starting from analytical considerations; a first modellization using Mode-Matching (MM) tools was applied to the reference horns; the Elegant BreadBoard (EBB)  4KRL model was built and the model results verified; a Finite Element Method modellization  (FEM) was applied to the horn-target system, allowing to refine the design, which was verified in the calibration test on the Flight Model (FM) parts.

\begin{figure}
		\includegraphics[width=\linewidth]{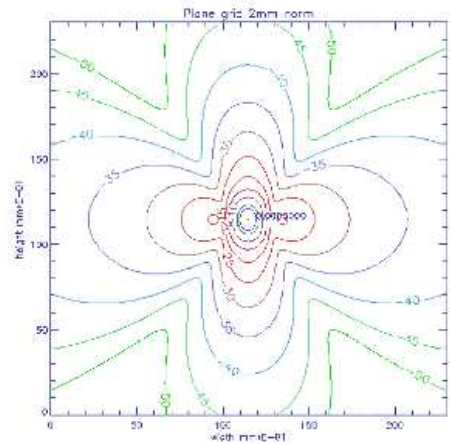}
		
	\caption{Near-field contour plot of a 4KRL pyramidal horn. The pattern is cross-shaped and targets are designed to match it.}
	\label{fig:horn pattern}
\end{figure}

Mode Matching (MM) modelling: The first optimization on the horns (\cite{2004NIMPA.520..396C}) was devoted to minimize the mismatch at the aperture. Horn dimensions were scaled from 'standard' pyramidal horns, but it resulted in {\it non standard} electromagnetic design, where flare length is strongly reduced, impacting mainly the mismatch at the aperture and the antenna power pattern shape. The near-field pattern was calculated using GRASP\footnote{
GRASP is a software developed by TICRA (Copenhagen, DK) }: a typical example is reported in Figure \ref{fig:horn pattern}. 

Each target is basically a rectangular ECCOSORB\texttrademark CR block, shaped for optimal matching with the incoming field. The back part is made of highly absorbing CR117, while the front sector, made of CR110, reduces the mismatch. A cross-shaped void is grooved in the front part, to match the pyramidal horn near field. A pyramid, made of CR110, is placed in the center, contributing both to absorb the peak of the incoming field and to reduce reflections by geometrically tapering the absorber (see Figure \ref{fig:target}).
Other target designs (such as bed-of-nails) have been studied and some prototypes built and tested. Yet, the selected one showed the best performance for target of such small dimensions.
At first, the design was linearly scaled with the frequency to all the LFI bands.
We found that the return loss at the horn aperture is the dominant term. 
Some considerations were drawn: while horn aperture dimensions linearly scale with the frequency, waveguides dimensions and ECCOSORB\texttrademark RF properties do not;  for mechanical reasons, the horn-target distance is the same at all frequencies. Since MM modelling is not able to study the spillover at horn-target gap, a FEM modedelling technique \citep{2004NIMPA.520..396C} was applied in the second optimizations step, where the horn-target combined system was separately optimized in each of the LFI bands. 

FEM analysis required an adequate knowledge of electromagnetic characteristics: dedicated measurements have been performed on materials (see results in the Appendix).

\begin{figure}
	\centering
		\includegraphics[width=\linewidth]{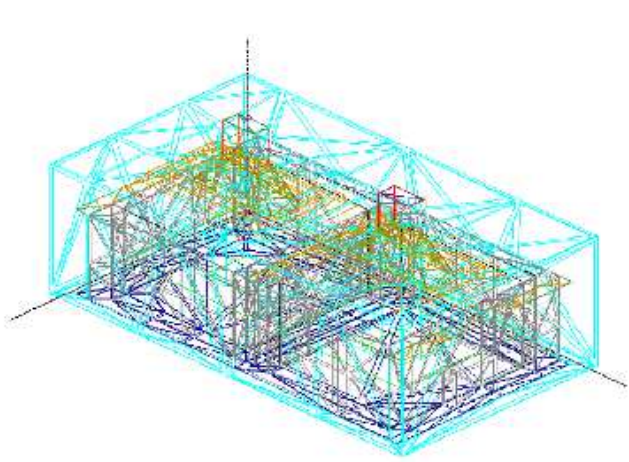}
	\caption{Graphics representation of the FEM discretization in	thetraedra: two	horns and	two targets are	modelled together. The environment radiation box is represented in light blue}
	\label{fig:FEM_modelling}
\end{figure}

The FEM conceptual scheme is reported in Figure \ref{fig:FEM_modelling}, where ports are placed at the waveguides ends. TE10 mode is fed to the waveguides and it is possible to evaluate the S-parameters: the Sii parameter (Reflection) together with the Sij  parameter (Cross Talk radiation coming from the port i detected from the port j when two coupled horns are modelled).

\begin{figure}
	\centering
		\includegraphics[width=\linewidth]{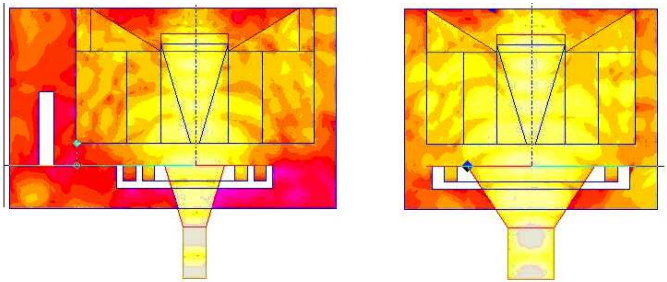}
	\caption{E-field distribution for the 70 GHz FM Model (left panel: PHI= 90 cut; right panel: PHI=0 cut); lighter colours correspond to stronger E field. The 
white colour corresponds to areas (perfect conductors)  where radiation does not enter. The metal case around  the target  is here replaced by a perfect conductor 
boundary  condition. The metal  baffle is evidenced (white) in the left panel.}
	\label{fig:target_field}
\end{figure}

A radiation box, surrounded by perfectly matched layers, is placed around the system to evaluate the radiated field. In this way the spillover can be evaluated. 
Reference waveguides, with the actual complex routing, were separately modelled. 

Finally, an accurate analysis was performed to investigate the stray-light contribution from sky external sources. 

The final results of this long modelling process is a different design, both for horns (with waveguides) and loads, for each LFI band.

The final design for the horns, in addition to changes in the overall dimensions (flare angle and mouth size), is innovative in the introduction of rectangular $\lambda /4$ grooves, around the aperture (Figures \ref{fig:70GHz} and \ref{fig:4K_WG_lowf}), able to significantly reduce the spillover (acting as rectangular waveguide chokes). Due to mechanical constraints, one complete groove and one on the long side (E-plane) are machined on each horn.  70 GHz RH have also been provided with a special small baffle, protruding 1.5 mm from Front End Modules, partially shielding the horn-target gap, therefore contributing in reducing the leaking radiation. 	

\begin{figure}
	\centering
		\includegraphics[width=\linewidth]{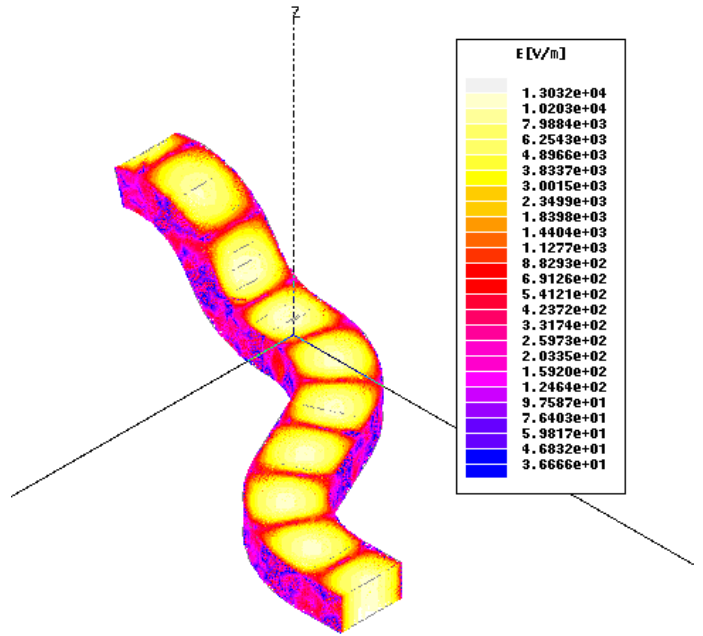}
	\caption{Reference waveguide 28L modelled with FEM: the E-field distribution is represented with a logarithmic temperature colour palette)}
	\label{fig:wg_fields}
\end{figure}

The final target design  for the 44 and 30 GHz bands includes only small modification with respect to the original one: front dimensions in the E-field direction was increased and the thickness of the back CR117 layer was modified. Targets for the 70 GHz were deeply modified by FEM modelling results.
The main differences are: the back ECR117 part is shaped as a pyramid, topped with a complementary part made of ECR110; the tip of the central pyramid is protruding 1 mm out of the front target surface. The final dimensions are reported in Table \ref{table:rt_data}.

\begin{table}[h!]
\centering
	\begin{tabular}{lccccccc}
	\hline\hline
Frequency	&A1 & A2 &	B &	C &	D &	E 	& F\\
\hline
GHz & \multicolumn{5}{c}{mm}\\
	\hline
70 	& 16.29	& 14.29 	& 3.30	& 1.89	& 4.63 	& 6.29 & 	2.32\\
44 	& 24.73 	& 22.73	& 5.00	& 3.00	& 7.36	& 10.00 &	3.69 \\
30  	& 33.34 	& 33.34	& 5.00	& 4.40	& 10.79	& 14.67 & 	5.39\\
	\hline

\end{tabular}
\caption{Reference Targets design dimensions. Refer to Figure \protect{\ref{fig:rt_dim}} for the meaning of each column.}	
\label{table:rt_data}
\end{table}

\begin{figure}
	\centering
	\includegraphics[width=\linewidth]{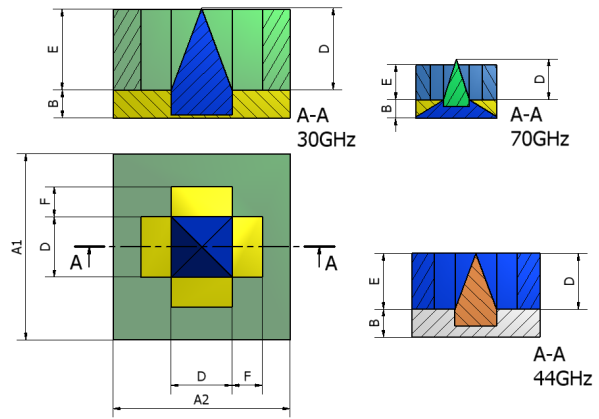}
	\caption{Sketch of the reference target. Letters refer to Table \protect{\ref{table:rt_data}}. Note that the yellow part (base) is separated in two sub-parts at 70 GHz.}
	\label{fig:rt_dim}
\end{figure}

\begin{table*}
	\centering
		\begin{tabular}{lccccccccc}
	\hline
Frequency &WG&	c& d&	e&	f&	g&	g1&	g2&	g3\\
\hline
GHz & & \multicolumn{2}{c}{deg} & \multicolumn{6}{c}{mm}\\	
\hline
70	& WR12	& 33.58	& 16.50	& 4.08	& 8.77	& 4.27	& 0.43	& 0.85	& 1.13\\
44	& WR22	& 34.08	& 10.68	& 5.36	& 14.73	& 6.68	& 0.68	& 1.36	& 1.80\\
30	& WR28	& 47.57	& 28.68	& 10.80	& 21.60	& 6.62	& 1.00	& 1.99	& 2.64\\
	\hline
	\end{tabular}
\caption{Reference Horns dimensions. Refer to Figure \protect\ref{fig:rh_dim} for the meaning of each column.}	
\label{table:rh_data}
\end{table*}

Moreover, mechanical contraints, imposed minor further modifications to targets, to avoid interference between parts and  to allow their positioning in the final system. The impact of such changes were measured during the characterization campaign and resulted to be negligible. 

\begin{figure}
	\centering
	\includegraphics[width=\linewidth]{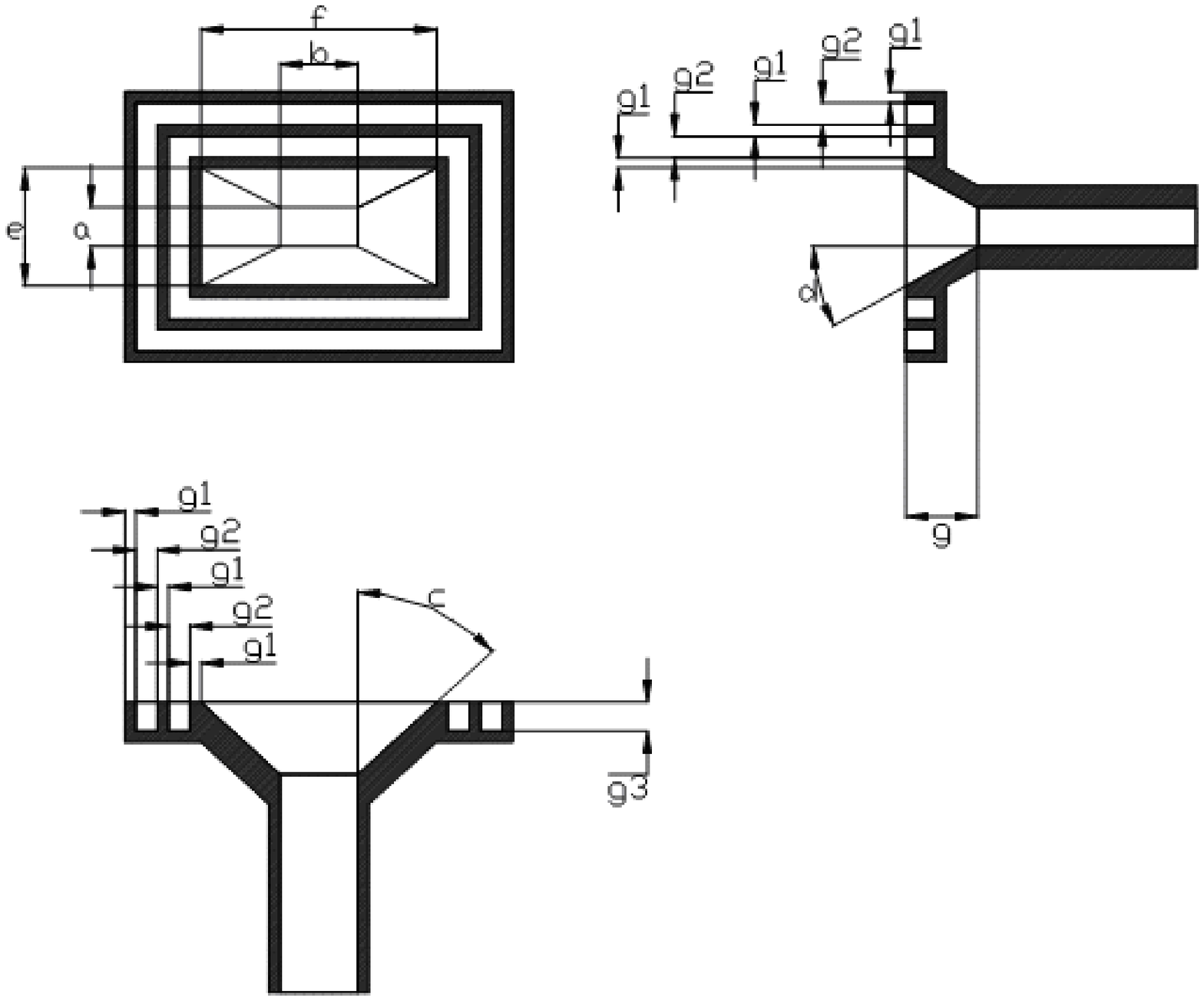}
	\caption{Schematic view of a Reference Horn. Letters refer to Table \protect{\ref{table:rh_data}}}.
	\label{fig:rh_dim}
\end{figure}

\subsection{Thermal design}

The 4KRL thermal design was studied to simplify as much as possible the interfaces. Thermal
interface is dominated by conduction through thermal washers (the only point of contact between the RT and the HFI shield), allowing careful control of the heat transfer. Contact surfaces are mirror-polished at sub-$\mu$m level. The method of bonding RT parts, which leave the base as the only thermal link, allows heat flow to be modelled in one dimension.
Metal parts are assembled using Stainless Steel screws at high torque, to make thermal contact as close as possible to an ideal value.

Temperature of the 4KRL will be measured in flight by temperature sensors mounted inside the HFI shield. High sensitivity sensors are placed close to the 4KRL interface, both on the upper conical part (where 70GHz loads are) and on the lower, cylindrical one (30 and 44 GHz loads). Sensitivity is of the order of 0.1 $\mu$K and accuracy at the level of mK \citep{Lamarre09}.

\begin{figure}[ht!]
	\centerline{\epsfig{file=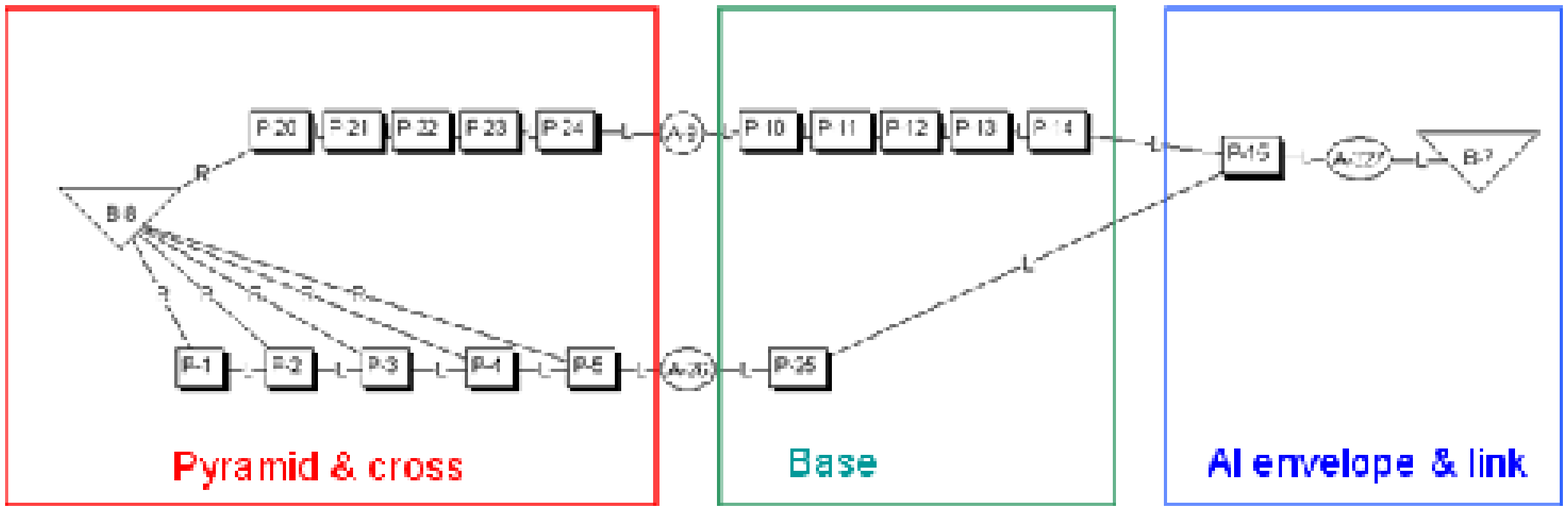,width=9.cm}}
\caption{The scheme of the thermal model for one of the 30 GHz reference targets. The Aluminum envelope is condensed in one node linked to the HFI shield boundary node. The pyramid and the cross follow two parallel paths, coherently with heat flowing in the axial direction. The base of the pyramid is condensed in one node, while the base of the cross is split in five slices. Both pyramid and upper cross side are divided in five nodes each.}
\label{fig:thermalmodel}
\end{figure}

A thermal model of the 4KRL unit was realised. We started from the general design of the single reference target (see Figure \ref{fig:target}).
Each part of the target, the base, the cross and the pyramid is divided in five layers orthogonal to the pyramid axis. The aluminum case where the ECCOSORB\texttrademark parts are bonded and the support structure for the 70GHz loads are assumed as perfect thermal conductors, due to their small thickness and mass. 
The optimization of thermal washers allowed to increase the damping factor from the QM to the FM unit.
The thermal model network representation is shown in fig. \ref{fig:thermalmodel}.

The main approximations of the current model are:

\begin{itemize}
\item due to the very small distance (1.5 mm) the radiative link is just considered with a perfect view factor, no analytic detailed radiative model was implemented
\item Eccosorb CR110 thermal conductivity and specific heat are used for all ECCOSORB\texttrademark nodes  
\end{itemize}

\begin{figure}[ht!]
	\centerline{\epsfig{file=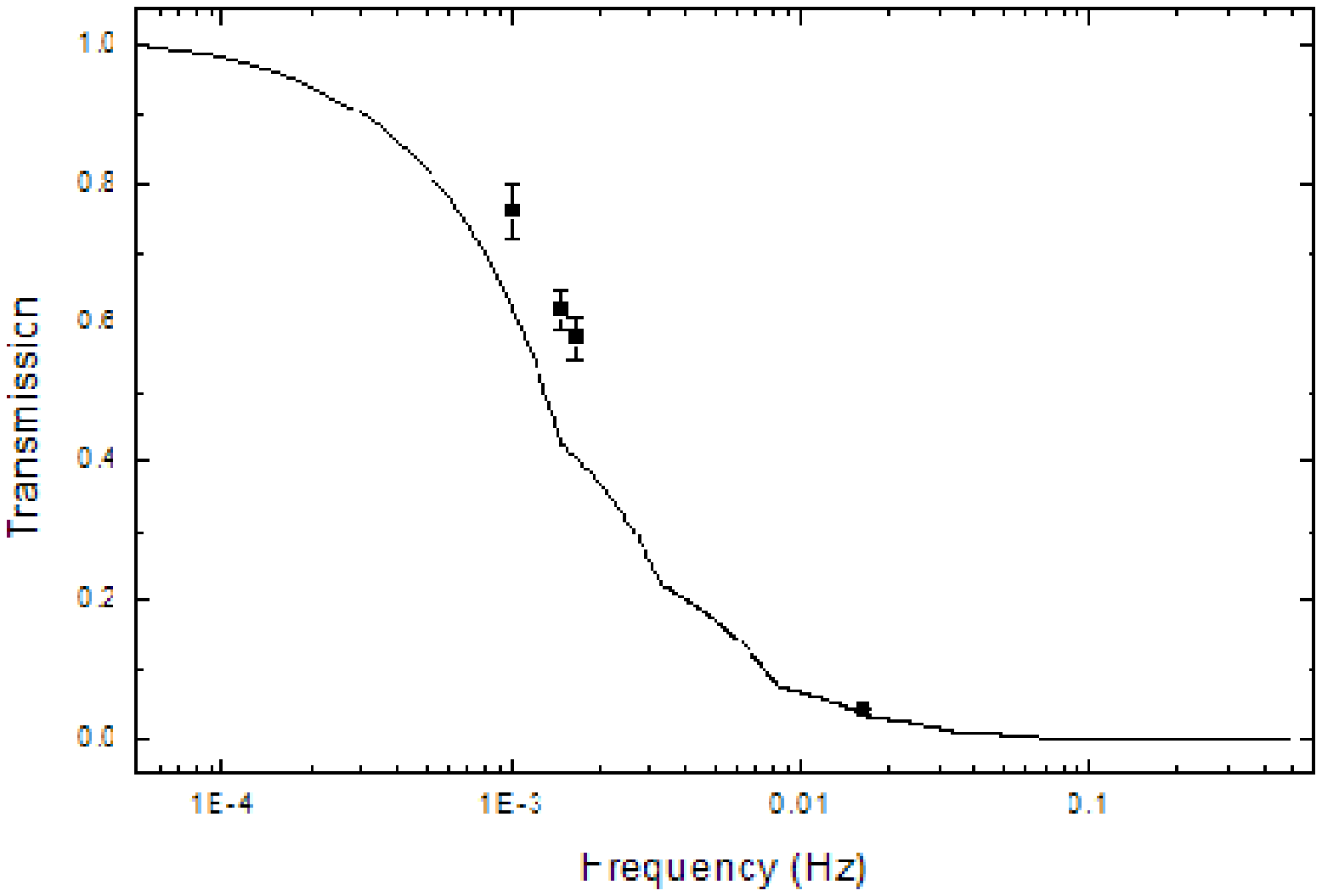,width=9.cm}}
\caption{Comparison between the simulated transfer functon foir a 44 GHz reference target (line) and the measured data. }
\label{model}
\end{figure}

The second point above is evidenced when comparing, for instance, the transfer functions of thermal fluctuations with measured data.
Figure \ref{model} shows how the measured values suggests a higher cut frequency with respect to what is foreseen by the model. In any case, the agreement is satisfactory and it shows the level of model prediction of the loads dynamic behavior.

\subsection{Mechanical Design, Properties and Manufacturing }

\begin{figure}[htbp]
	\centering
		\includegraphics[width=\linewidth]{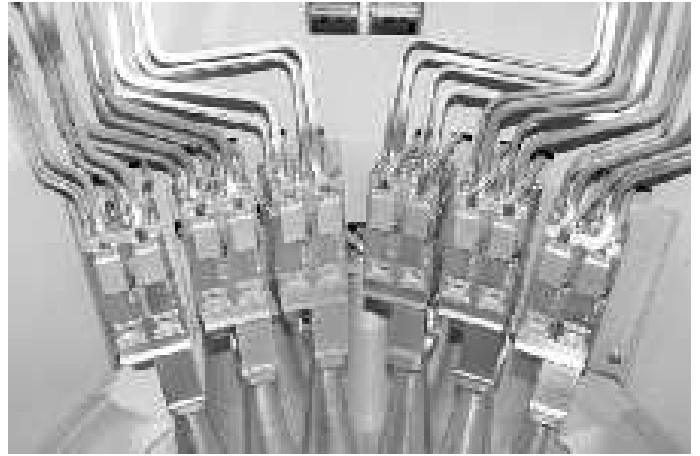}
	\caption{70 GHz radiometers mounted on the LFI Main Frame structure. Reference Horns are clearly visible. Note also the step at the level of the OMT, used to cover the horn-target gap and reduce the spillover.}
	\label{fig:4K_WG_70GHz}
\end{figure}

\begin{figure}[htbp]
	\centering
		\includegraphics[width=\linewidth]{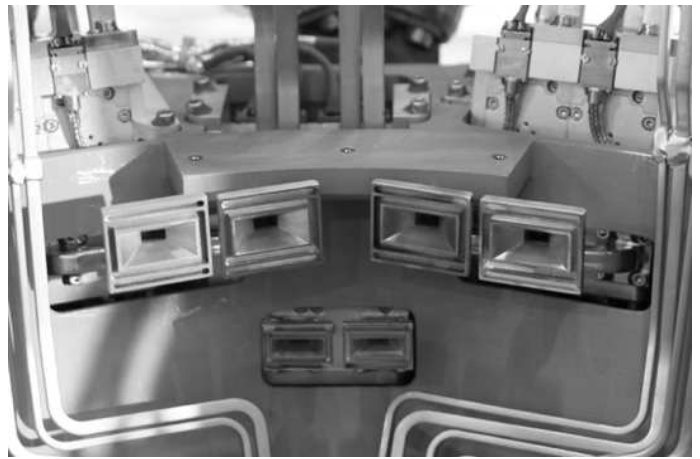}
	\caption{30 GHz reference horns (upper, larger ones) and two of the 44 GHz ones (lower) and their reference waveguides mounted on the LFI Main Frame structure. The grooves around the aperture are clearly visible.}
	\label{fig:4K_WG_lowf}
\end{figure}

The location of the LFI radiometers on the FPU (see Figure \ref{fig:mainframe}) lead to separate the loads in two different 'sectors': one for the 70GHz targets, forming the 4KRL upper sector, mounted on the cone supporting the HFI feed-horns; one for the 30 and 44 GHz targets, the so-called 4KRL lower sector, mounted on the cylindrical part of the HFI 4K shield. The 4KRL lower sector is then composed by the 30 GHz target assembly (largest targets at the bottom of Figure \ref{fig:4K_montato}) and three identical parts, one for each of the LFI 44 GHz FEM, mounted at 120$^{\circ}$ from each other (one of them is shown in Figure \ref{fig:4K_montato} over the 30 GHz ones).
Details in the mechanical design have been optimized to allow the integration of the unit on the HFI completely assembled. The final result is shown in Figure \ref{fig:4K_montato}. While RH for the 70GHz radiometers were obtained inside the Front-End Modules (Figure \ref{fig:4K_WG_70GHz}) , the location of the 30 and 44GHz FEMs required the introduction of external waveguides (Figure \ref{fig:4K_WG_lowf}).

Each reference target is an assembly of 3 or 4 (at 70GHz) parts obtained by casting ECCOSORB\texttrademark. This material is an epoxy resin, loaded with iron particles with a typical dimension of few $\mu$m (see image in the Appendix). CR-110 differs from CR-117 in relative density of iron particles. The mould used was built in a flexible material (Derlin)  in order to allow an easier removal operation and to avoid crack generation during the cooling process after the polymerization.
Only the cross parts are machined to create the final shape. The required tolerance is achieved by an optimal control of the casting process, in particular for the thermal contraction derived from the polymerization temperature of 80C and from the high thermal contraction coefficient of the absorber. The process was studied to avoid the presence of air-bubbles inside the parts, a vacuum die-casting process was set to achieve this requirement. The curing time was enough fast to avoid iron particles to settle, avoiding the use of Cab-o-sil powder as in other applications \citep{Hemmati, 1999ApJ..512..511M}. However, some parts were manufactured using the Cab-o-sil, and reflectance was measured. No significant difference was found. Some parts were also cut in order to verify the presence of bubbles and X-ray inspection was done in order to verify the quality of the process. A very careful check of the surface was done by using a microscope for all the components.

\begin{figure}
	\centerline{\epsfig{file=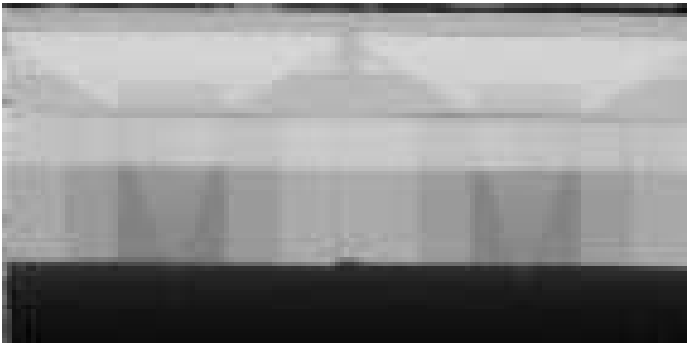,width=9.cm}}
	\caption{X-ray image of one of the 70GHz reference target. Different ECCOSORB\texttrademark parts are clearly visible.}
\label{fig:RT_Xray}
\end{figure}

ECCOSORB\texttrademark targets are assembled and then bonded, using Hysol EA9394\footnote{Hysol is a Trade Mark of Loctite Aerospace.} inside aluminum enclosures, made of certified Al6061-T6 to ensure optimal thermal properties at low temperature (see the Appendix for thermal data). Adhesive is not placed on target lateral surface, reducing mechanical stress induced by differential thermal expansion coefficients between ECCOSORB\texttrademark targets and Al cases.
To increase safety, in the unfortunate case of a failure in the adhesive, ECCOSORB \texttrademark parts are also mechanically locked within the metal enclosure. Alignment pins ensure the correct alignment of the whole unit within 0.1 mm.


Stainless steel (AISI304) thermal washers are interposed between the loads and the interface point to the HFI. These are small cylinders (typically 5 mm long, 1 mm wall thickness) whose dimension are optimized to dump temperature fluctuations in order to meet the requirements. 

\begin{figure}[ht]
	\centering
		\includegraphics[width=\linewidth]{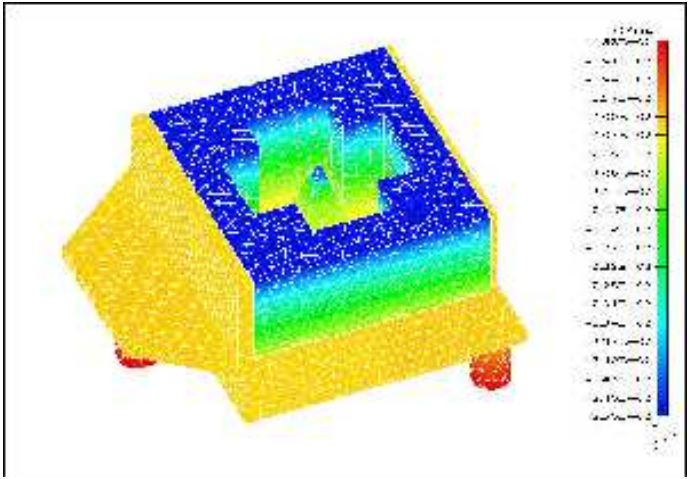}
	\caption{FEM analysis results of relative temperature distribution on reference targets, used for thermo-structural analysis.}
	\label{fig:RT_thermal}
\end{figure}

\begin{figure}[ht]
	\centering
		\includegraphics[width=\linewidth]{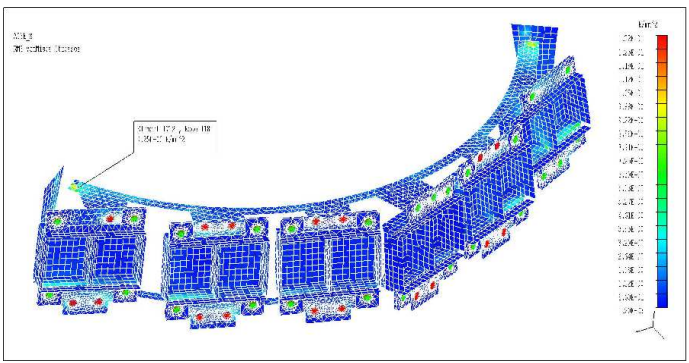}
	\caption{Random vibration analysis of the 70GHz loads. The most critical areas, as expected, are the fixation points between the supporting structure and the lateral brackets.}
	\label{fig:RT_mechanical}
\end{figure}

\begin{figure}[ht]
	\centering
		\includegraphics[width=\linewidth]{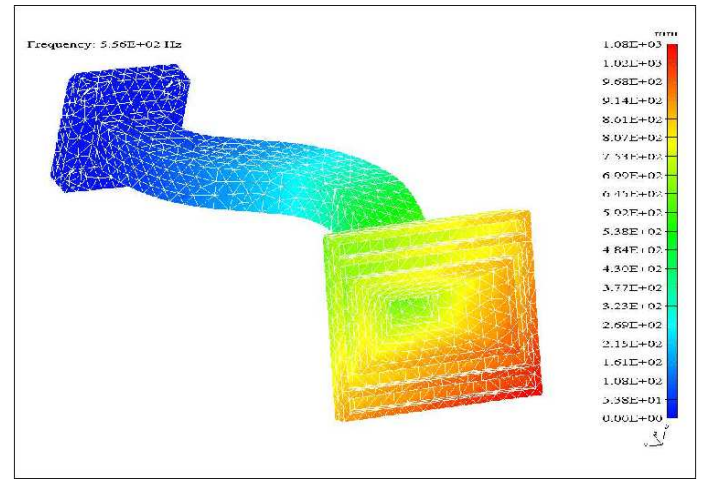}
	\caption{Example of FEM analysis results on reference horns. It was used to determine the first natural frequency in terms of displacement.}
	\label{fig:RW_thermal}
\end{figure}

ECCOSORB\texttrademark elastic module was measured on a representative sample, produced using the qualified process (see the Appendix).
A FEM analysis was performed in order to verify the mechanical stress induced by the thermal contraction (see an example in Figure \ref{fig:RT_thermal}). It was confirmed that the most critical area is around the base of the target. In order to verify the capability of the reference target to tolerate the high vibrations of the launch, a FEM model of all the 4KRL was developed: the natural frequency was calculated and a random vibration analysis was performed on the modal model in order to verify the effect of the vibration test level on the structure (Figures \ref{fig:RT_mechanical} and \ref{fig:RW_thermal}). 

Reference waveguides, used at 30 and 44 GHz, are characterized by a complex routing. Two or three bends are needed in a length less than 10 cm. They are obtained by electro-forming pure copper on Al mandrel, which is then solved in caustic soda. In order to achieve the required dimensional tolerance, mandrels are electro-eroded, obtaining a final measured accuracy of about 0.05 mm.
Copper reference horns and gold-plated brass flanges, machined separately, are then soft-soldered. A small layer of gold is then flashed on the external surface to avoid oxidation. 
A first step in the qualification process was to determine the properties of the electroformed copper used for the waveguide, defining a test on specimens representative of a 30 GHz waveguide (see data in the Appendix). Then, a FEM model of the waveguide was produced to determine the stress level and to determine the safety margin of the part. 

\begin{figure}[ht]
	\centering
		\includegraphics[width=\linewidth]{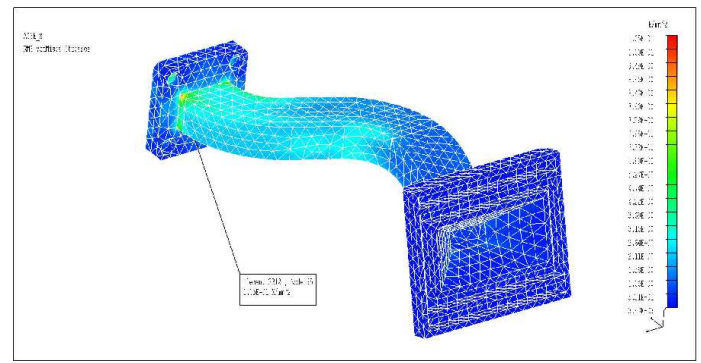}
	\caption{Example of random vibration analysis on one of the 30GHz reference horns. The most critical area, as expected, is the interface between the waveguide and its flange.}
	\label{fig:RW_mechanical}
\end{figure}

The mass of the FM 4KRL unit is 572 g for the targets, including screws (mounted on the HFI) and 306 g for the waveguides (mounted on the LFI).

\section{Test activity\label{section:performance}}

The 4KRL unit model philosophy comprises four different models. Prototype Demonstrators (PD) and Elegant BreadBoard (EBB) models were used to design both horns and targets. Some loads were also used in testing the LFI radiometers prototypes. This activity allowed us to complete the 4KRL design. The 
Qualification Model (QM) was manufactured and submitted to a complete set of tests. Reference targets and horns were vibrated and successfully met the mechanical requirements. The RF and thermal test results were used to further refine the design (i.e. thermal dumping was increased, mounting structure was slightly modified to facilitate the integration), resulting in the Flight Model 4KRL design. The FM was then submitted to a complete characterization and acceptance test campaign, where RF, thermal and mechanical properties were fully measured. 

\subsection{RF performance}
\begin{figure}
	\centering
		\includegraphics[width=\linewidth]{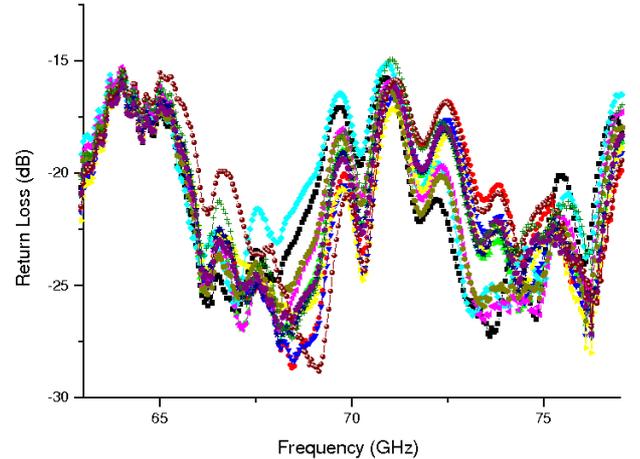}
	\caption{Return Loss for the 70 GHz targets coupled with a representative reference horn. To be noted the remarkable uniformity among all loads, due to a robust design and manufacturing.}
	\label{fig:RL70}
	\end{figure}

\begin{figure}
	\centering
		\includegraphics[width=\linewidth]{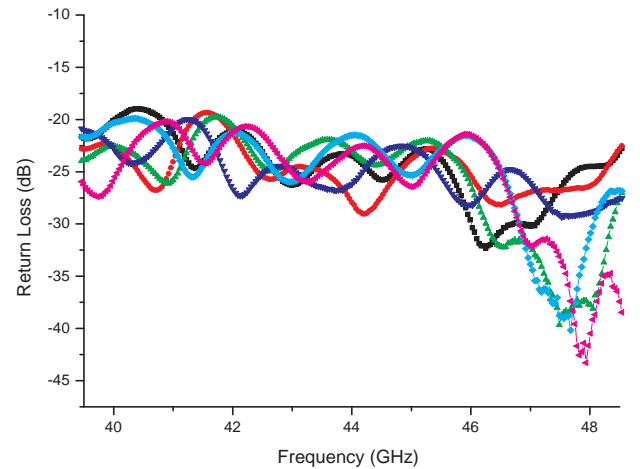}

	\caption{Return Loss for the 44 GHz targets coupled with their reference horns.}
	\label{fig:RL44}
\end{figure}

\begin{figure}
	\centering
		\includegraphics[width=\linewidth]{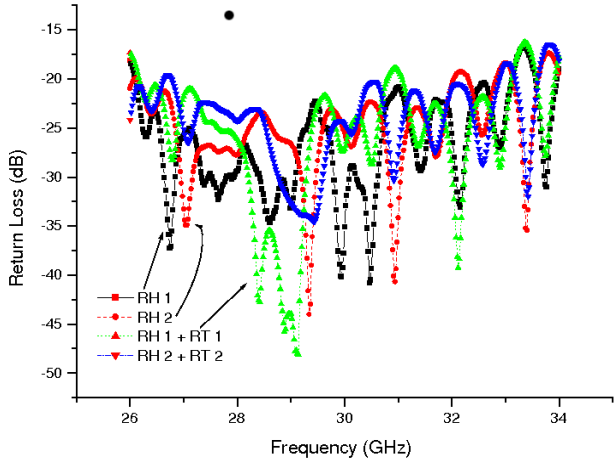}

	\caption{30 GHz loads. The plots reports the RL measured for the horns in anechoic environment and coupled with their reference loads.}
	\label{fig:RL30}
\end{figure}

The 4KRL parts have been tested for RF performance before assembling into the instrument Focal Plane Unit (FPU). Each target was mounted on a 3-axis positioning system, with micrometric control. A Scalar Network Analyser (SNA) was used to measure the RF performance, connected to the reference horns. While FM reference horns were used at 30 and 44 GHz, representative horn with adaptive flanges were adopted for the 70 GHz loads, where the real horns are integrated into the radiometer Front-End Modules. Return Loss (RL), Insertion Loss (IL) and cross-talk between adiacent loads was measured for all targets. The results are reported in Tables \ref{table:rh_rf} and Figures \ref{fig:RL70}, \ref{fig:RL44} and \ref{fig:RL30}. 

\begin{table}[htdp]

\begin{tabular}{lccc}

LFI RCA & Average IL (dB) &\multicolumn{2}{c}{Average RL (dB)}\\
& RH  & RH only & RH + RT\\
\hline
\hline
18 -M & 0.16 & -20.99 & -20.14\\
18 - S  & - & - & -20.18\\
19 - M  & - & - & -20.54\\
19 - S  & - & - & -20.25\\
20 - M  & - & - & - 20.18\\
20 - S & - & - & -19.56\\
21 - M  & - & - & -20.58\\
21 - S  & - & - & -20.48\\
22 - M  & - & - & -20.07\\
22 - S  & - & - & -20.29\\
23 - M  & - & - & -19.57\\
23 - S  & - & - & -19.83\\
\hline
24 - M & 0.09 & -24.13 & -24.21\\
24 - S & 0.08 & -24.20 & -23.42\\
25 - M & 0.08 & -22.39 & -23.92\\
25 - S & 0.11 & -22.27 & -24.04\\
26 - M & 0.09 & -22.27 & -23.85\\
26 - S & 0.08 & -21.01 & -23.22\\
\hline
27 - M & 0.11 & -26.59 & -24.49 $^*$\\
27 - S & 0.10 & -25.15 & - 26.45 $^*$\\
28 - M & 0.09 & -24.48 & -24.49\\
28 - S & 0.10 & -25.78 & -23.92\\
\hline
\end{tabular}
\caption{RF measured performance for the 4KRL: RH (+WaveGuide) Insertion Loss, RH (+WG) RL, RH (+WG) + RT Return Loss. 70 GHz performance have been measured using a representative Reference Horn and waveguide, since RHs are internal to FEMs: this value is reported in the Table. RCAs 24-26 are the LFI 44 GHz channels, 27 and 28 the 30 GHz ones. M and S refer to Main and Side OMT arm, respectively. $^*$ measured on the FM RTs and a representative Reference Horn. }
\label{table:rh_rf}
\end{table}

We note a substantial uniformity of RL data at 70 GHz. Some of the loads, LFI 23 (M and S) and LFI 20 - S are close to the requirements of an average RL lower than -20 dB at all the LFI frequencies. This is considered a minor effect, due to the specific geometry of those reference targets, and the parts have been accepted for flight. 30 and 44 GHz loads show an overall RL well within the requirement. Minor differences in the measured values are present, since targets are not identical among each other: due to the orientation of the LFI FEMs, the shape some of the 70 GHz target were adapted by cutting a small portion on the back of the ECCOSORB\texttrademark  part; the central targets of the 30 GHz assembly are also been adapted in the design, due to mechanical constraints; differences due to bonding can also be present. 
In Figure \ref{fig:RL30} we compare, as an example of the typical behavior, the results of reference horns matched to their targets (RH1+RT1, RH2+RT2) and to an ideal absorber (RH1, RH2). We observed that the RL performance is mainly dominated by the reference horn design. Measured performance is in good agreement with the FEM RF model.
The effect of target-horn misalignment was also measured. The effect of $\pm 1$ mm linear displacement on the three axis and of $\pm 1^{\circ} $ around them resulted in a maximum effect of $ 1$ dB for all the LFI bands.


The 70GHz loads are the most affected by the spillover, because of their location close to the top of the Focal Plane Unit. To evaluate the SPO quantity, a dedicated detailed modelling was implemented, considering the whole cavity and the contributions from the external environment. It resulted in a gobal straylight rejection ($SPO_{eq}$) better than -60 dB, well within the required limit.

\begin{figure}
	\centering
		\includegraphics[width=\linewidth]{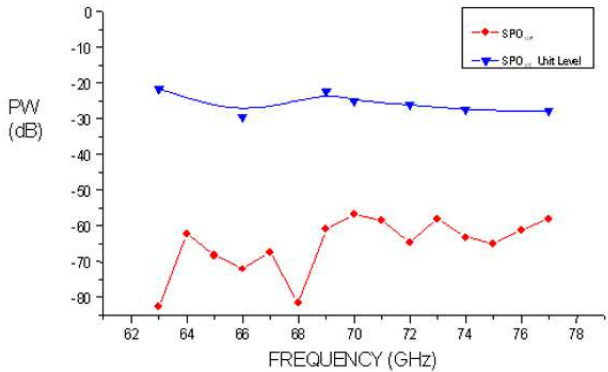}

	\caption{Results of the spillover modelling for the 70GHz loads. Triangles refer to the internal spillover, diamonds to the global straylight rejection.}	
\label{fig:SPO_model}
\end{figure}

\begin{figure}
	\centering
		\includegraphics[width=\linewidth]{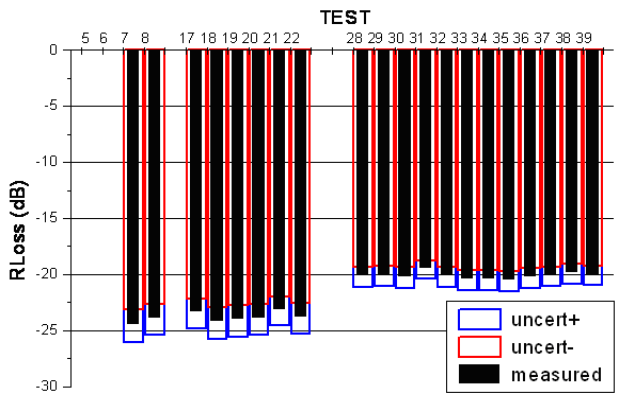}

	\caption{Comparison of average RL measured on the LFI RTs coupled with their RHs. Tests 7-8 refer to the 30 GHz loads, tests 17-22 to the 44 GHz, 28-40 to the 70 GHz loads. Measurement systematic uncertainties, mainly due to the measurement setup, are also reported. The required values is RL $\le$ -20 dB.}	
\label{fig:rl_comp}
\end{figure}

\begin{figure}
	\centering
		\includegraphics[width=\linewidth]{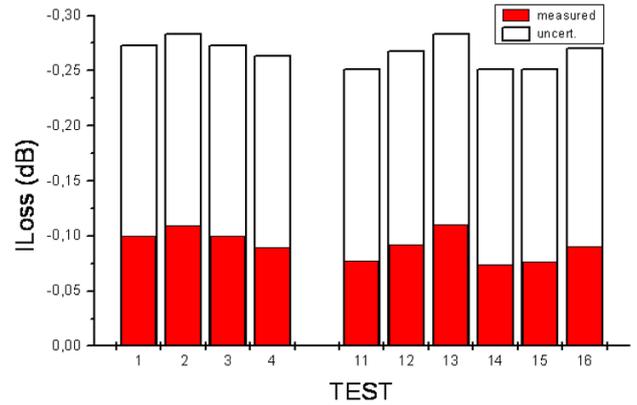}

	\caption{Comparison of IL measured on the LFI reference wavegudes. Test 1-4 refer to the 30 GHz parts, tests 11-16 to the 44 GHz waveguides. Measurement systematic uncertainities are also reported.}
	\label{fig:IL_comparison}
\end{figure}

The IL of the 4KRL waveguides and horns, measured on the 30 and 44 GHz FM parts and averaged over the LFI bands, is reported in Figure \ref{fig:IL_comparison}.  While the sensitivity of the SNA is sufficient to measure IL of the order of 0.1 dB, the accuracy of the whole apparatus (mainly the directional coupler) leads to the large systematic uncertainty reported in the plot. The requirement of $IL_{4KRL} \le$ 0.15 dB is therefore considered met. Differences between the WGs are mainly due to differences in routing and length. The 70 GHz IL was modeled and compared with measure on the representative RH (see Figure \ref{fig:70GHz}), finding good agreement.

\subsection{Thermal performance}

Thermal tests were performed in the IASF-Bo 4K cryo facility, equipped with a GM cooler, with an heat lift up to 1.5 W at 4K. The 4KRL targets were mounted on a Al6061-T6 support, with mechanical interfaces representative of the HFI one. A temperature controlled Al cylinder was located around the loads, simulating the radiative thermal interface at 20K. This setup simulated the real environment in the payload, where targets are mounted on the HFI 4K shield in front of the quasi-cylindrical LFI main frame at about 20 K. It was also used to test the susceptivity to fluctuations of the LFI.

\begin{figure}[!h]
	\centering
 \subfigure[]{\epsfig{file=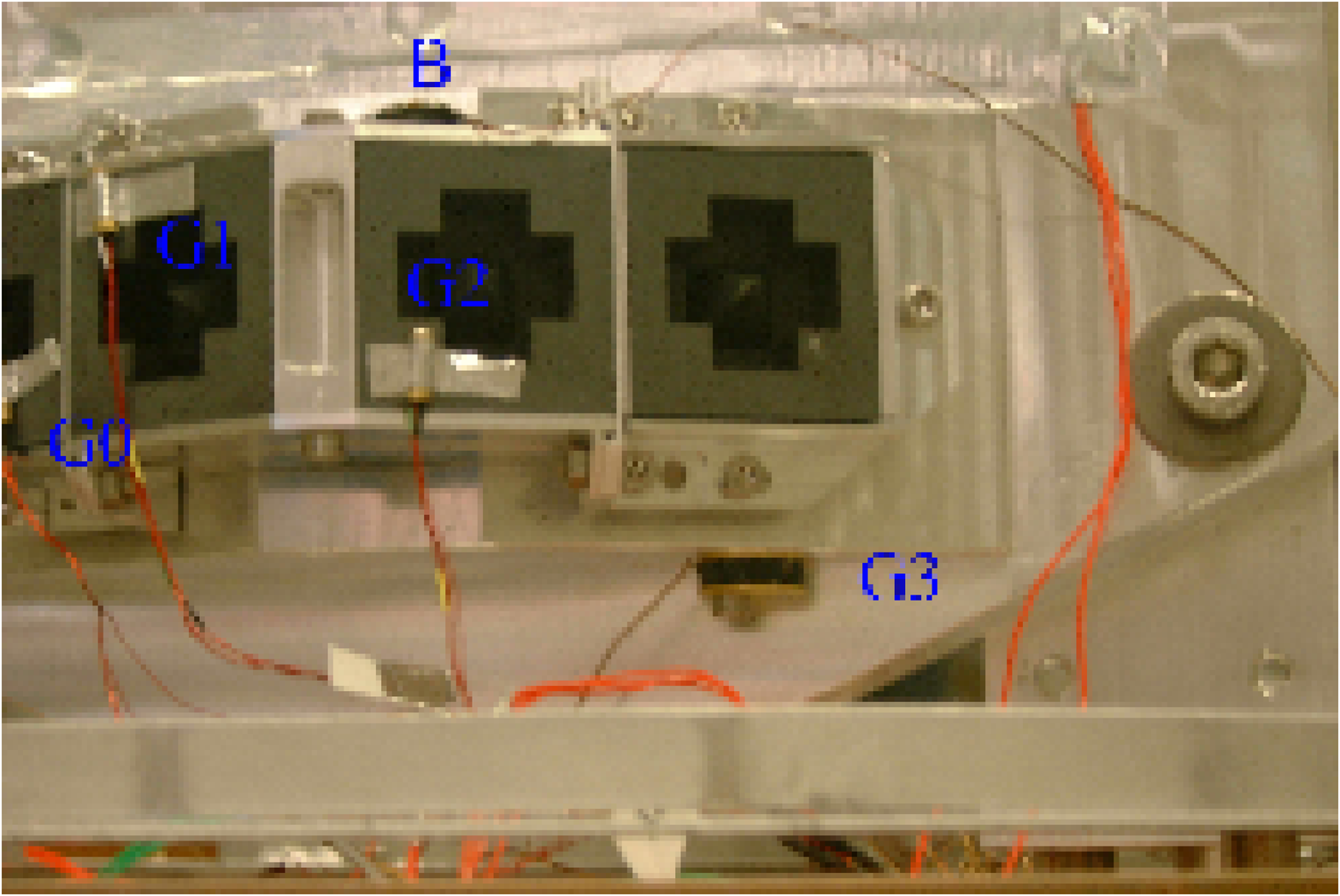,width=4.cm}} 
 \subfigure[]{\epsfig{file=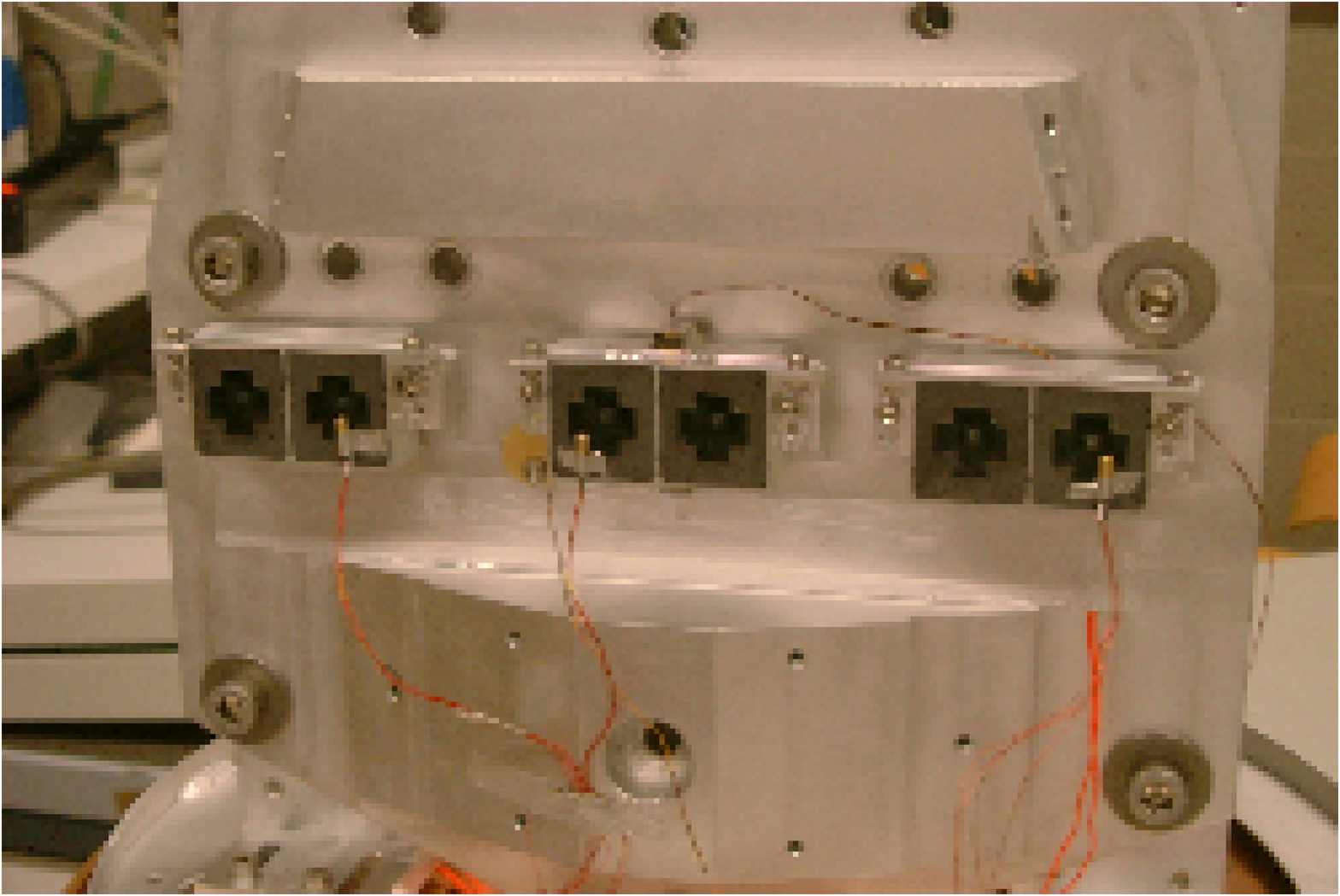,width=4.cm}}			
\caption{Sensors mounting on the 30 GHz (a) and 44 GHz (b) reference load targets, during thermal tests.}
\label{sens}
\end{figure}

Both the Al structures are connected to the cold flange via stainless steel thermal washers, whose dimensions are optimised to achive the desired stable temperatures with minimal heat load dissipated on the  cooler cold stage. The thermal regulation is obtained with Minco film heaters fixed to the shields.
Calibrated Lakeshore temperature sensors were used: silicon diode DT670, Cernox CX1050 and Germanium GR200A. Sensitivity and accuracy at 4 K, using a LakeShore temperature controller 340 readout, is better than 1mK and 30mK, respectively. The sensors are fixed with Aluminum tape on the face of the targets, as shown in Figure \ref{sens}.

The steady state measurements were used to estimate the radiative heat load coming from the 20 K shield to the reference targets.
The evaluation suffers of a large error because of the small temperature differences measured, comparable with the sensors accuracy.
The global heat load is evaluated as $(360 \pm 540)\mu W$, compliant with the requirement.

\begin{table*}
\centering
	\begin{tabular}{lcccc}
	\hline
Freq. & D (60s) &	D (600s) & D (667s) &	D (1000s)\\
	\hline
	30 GHz &	0.080 $\pm$ 0.004	&	0.60 $\pm$ 0.03	&	0.64 $\pm$ 0.03	&	0.78 $\pm$ 0.04	\\
	44 GHz &	0.133 $\pm$ 0.007   	&	0.81 $\pm$ 0.04	&	0.85 $\pm$ 0.04	&	0.91 $\pm$ 0.05 \\
	70 GHz &	0.131 $\pm$ 0.007   	&	0.72 $\pm$ 0.04	&	0.75 $\pm$ 0.04	&	0.85 $\pm$ 0.04 \\
	\hline
\end{tabular}
\caption{Thermal fluctuation damping measured for the reference loads at different frequencies}	
\label{th_data}
\end{table*}

\begin{figure}[ht!]
	\centerline{\epsfig{file=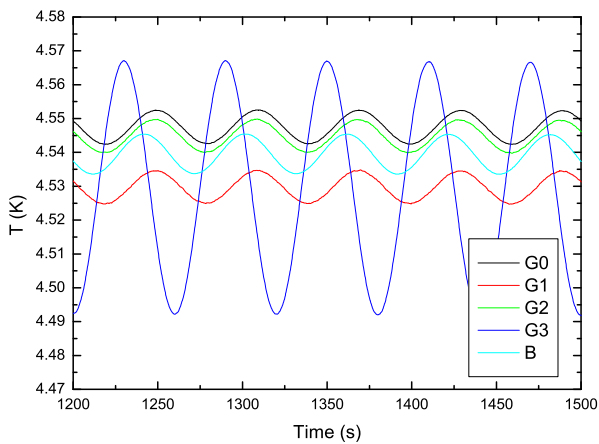,width=9.cm}}
	\caption{The temperature fluctuation damping measured during the 44 GHz thermal test. The 60 s fluctuation test data are shown. The blue line (B) is the temperature curve of the source of fluctuations while the damped curves refer to the aluminum case (G3) and target faces (G0-G2).}
\label{fluct}
\end{figure}

The thermo-mechanical damping was evaluated from the transient test, inducing sinusoidal temperature fluctuation with periods of 60, 600, 667 (typical Sorption Cooler period) and 1000 seconds at the level of the attachment point of the loads on the support structures. The fluctuation at the level of the targets is then acquired (Figure \ref{fluct}) and the transfer function (amplitude and phase) are estimated by the ratio of the amplitudes. The final results are summarized in the Table \ref{th_data}.

The damping factor for the 44 GHz and 70 GHz loads with 60 s period fluctuation was found slightly above the requirements. The effect was projected on simulated CMB  maps, and it was demonstrated that the final effect could be safely removed by destriping techniques \citep{2002A&A...387..356M}.

A set of additional functional tests consisting in thermal cycles of the load were performed to check the survival to thermo-mechanical stress.

\section{Conclusions\label{section:conclusions}}

The complete 4KRL unit was used for the LFI test activity at instrument level, where the LFI radiometer performance were measured in cryo environment. The unit was mounted on a dedicated support, thermally stabilized at a temperature around 20K, as well as the sky load. Performance of the LFI are then extrapolated at operating temperature. No temperature sensor was placed on the targets to measure the damping factor, but the global results on the 1/f noise performance was satisfying the performance of the LFI.  The effect of load temperature variation was also visible in the radiometer output and it was possible to remove it in data analysis. These results are fully described in \citep{Mennella_calibration}.

After this activity, the 4KRL unit was integrated on the HFI in October 2006. The final location was measured with respect to the design, using a 3-D measurement arm. The final alignment was found within a fraction of a millimeter with respect to the design values. After the integration of the HFI into the LFI, the clearance between reference horns and targets was verified using an endoscope. 

Once the unit is integrated on the payload, thermal stability of the RTs is directly linked to the HFI 4K cooler performance, while temperature data relevant to the 4KRL unit are provided by HFI sensors. These are mounted inside the HFI 4K shield and the temperature of the RTs is extrapolated using the thermal model described in a previous section. The 70 GHz loads, being located close to the HFI 4K flange, where the HFI sky horns, are mounted, will take advantage of the thermal stabilization obtained by a PID controller on the HFI internal shield. The 30 and 44 GHz loads, being located farther from the controlled stage, will be more influenced by the HFI 4K stages fluctuations. A full description of the HFI  cryo chain and its behavior can be found in \cite{Lamarre09}.

Preliminary results from ground based test at satellite level, performed at the Centre Spatiale de Lige (CSL), confirmed the excellent noise properties of the LFI radiometers, especially for those concerning the 1/f knee frequency, related to the 4KRL performance.  Therefore, indirectly, the correct behavior of the 4KRL unit was assessed (see \cite{Mennella_calibration} for details).

We have presented in this paper the design, development, manufacturing and test activities of the 4K Reference Load Unit for the Low Frequency Instrument on-board the Planck satellite. This work allowed to build an innovative, very compact, high performance cryogenic calibrator for millimeter wave radiometers, compliant with very stringent requirements. Relevant material data were collected from the literature and directly measured, where required. 

\begin{acknowledgements}
Planck is a project of the European Space Agency with instruments funded by ESA member states, and with special contributions from Denmark and NASA (USA). The Planck-LFI project is developed by an International Consortium lead by Italy and involving Canada, Finland, Germany, Norway, Spain, Switzerland, UK, USA. The Italian contribution to Planck is supported by the Italian Space Agency (ASI). The US Planck Project is supported by the NASA Science Mission Directorate.
We would like to thank ESA material division for the support given during test on materials. G. Dall'Oglio, L. Pizzo, Todd Gaier and Mike Seiffert helped us in the early stages of the 4KRL development. 
\end{acknowledgements}

\bibliographystyle{aa} 
\bibliography{valenziano_4K} 

\appendix

\section{Material data}

\begin{figure}[ht!]
	\centerline{\epsfig{file=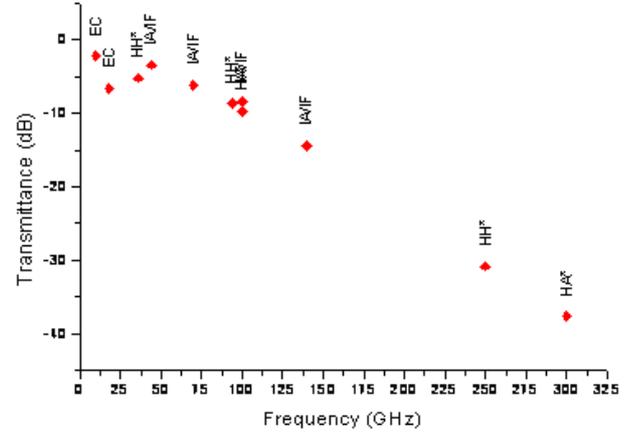,width=9.cm}}
	\caption{Logarithmic transmittance (expressed in dB) for ECCOSORB CR110. ECR: Emerson and Cumings, HH:\cite{Hemmati}, Ha: \cite{1986ApOpt..25..565H}, IA/IF: this work}
\label{fig:ECR_transm}
\end{figure}

\begin{figure}[ht!]
	\centerline{\epsfig{file=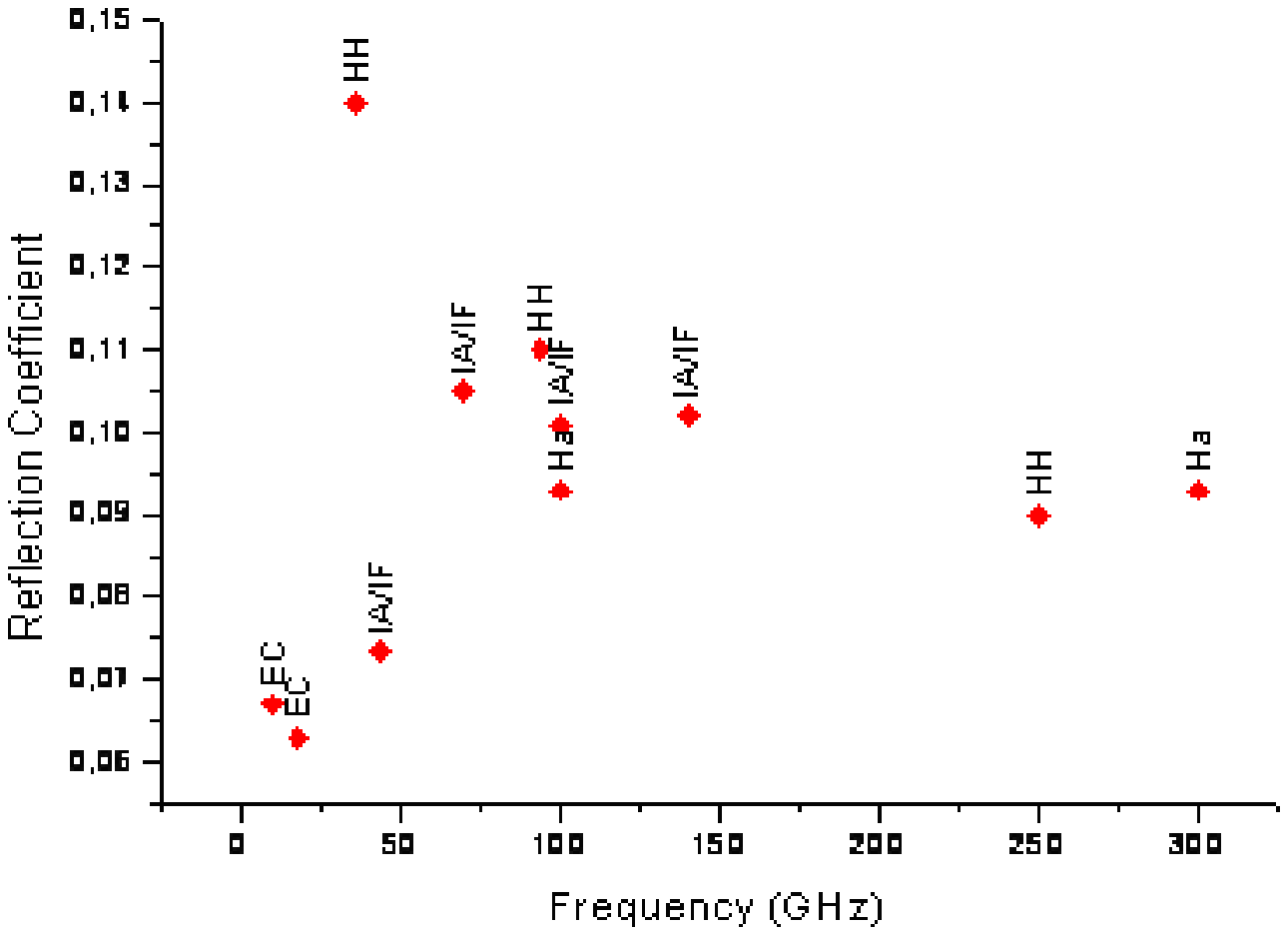,width=9.cm}}
	\caption{Reflection coefficient for ECCOSORB CR110. ECR: Emerson and Cumings, HH:\cite{Hemmati}, Ha: \cite{1986ApOpt..25..565H}, IA/IF: this work}
\label{fig:ECR_reflec}
\end{figure}

The 4KRL targets are made of ECCOSORB\texttrademark CR-series. This absorbing material, widely used in microwave passive components, has already been used in space applications and at low temperature \citep{1984IJIMW...5.1507P,1999ApJ..512..511M}. However, electrical and thermal properties are not well documented and it has been necessary to qualify it for our application. RF, thermal, mechanical and outgassing properties of Eccosorb CR-110 and CR-117 were measured at room temperature. RF transmittance and reflectivity data are reported in Figures \ref{fig:ECR_transm} and \ref{fig:ECR_reflec}. 

We report here some thermal and mechanical properties of the materials used in the 4KRL unit. Outgassing test, on samples prepared by the LFI team, were conducted by ESA.

\begin{table}
	\caption{Aluminium 6061 -  T6 Thermal properties at 4K (from CrioComp, Eckels Eng.)}
	\centering
	\begin{tabular}{lc}
\hline\hline
		Thermal conductivity W/(m K) & 9.53 \\
  		Specific heat  J/(Kg K)&  0.28\\
\hline
\end{tabular}
\end{table}	

\begin{table}
	\caption{Electro-formed copper mechanical properties.}
	\centering
	\begin{tabular}{lcc}
\hline\hline
		Elongation (\%) 	&  42.1 & 	this work \\
		Elastic Modulus (GPa) 	& 115.1 &	this work \\
		Yield Strength (MPa) & 168& 	this work\\
		Ultimate Strength (MPa) & 269.4 & this work \\
\hline
\end{tabular}
\end{table}	

\begin{table}
	\caption{Hysol EA 9394 Outgassing characteristics}
	\centering
	\begin{tabular}{lcc}
\hline\hline
			TML (\%) & 1.06 & \cite{ESA_outgassing_1} \\
			RML (\%) &0.34 &   \cite{ESA_outgassing_1}  \\
			CVCM (\%) &0.01  &  \cite{ESA_outgassing_1}  \\
		\hline
		\end{tabular}
\end{table}

\begin{figure}[htbp]
	\centerline{\epsfig{file=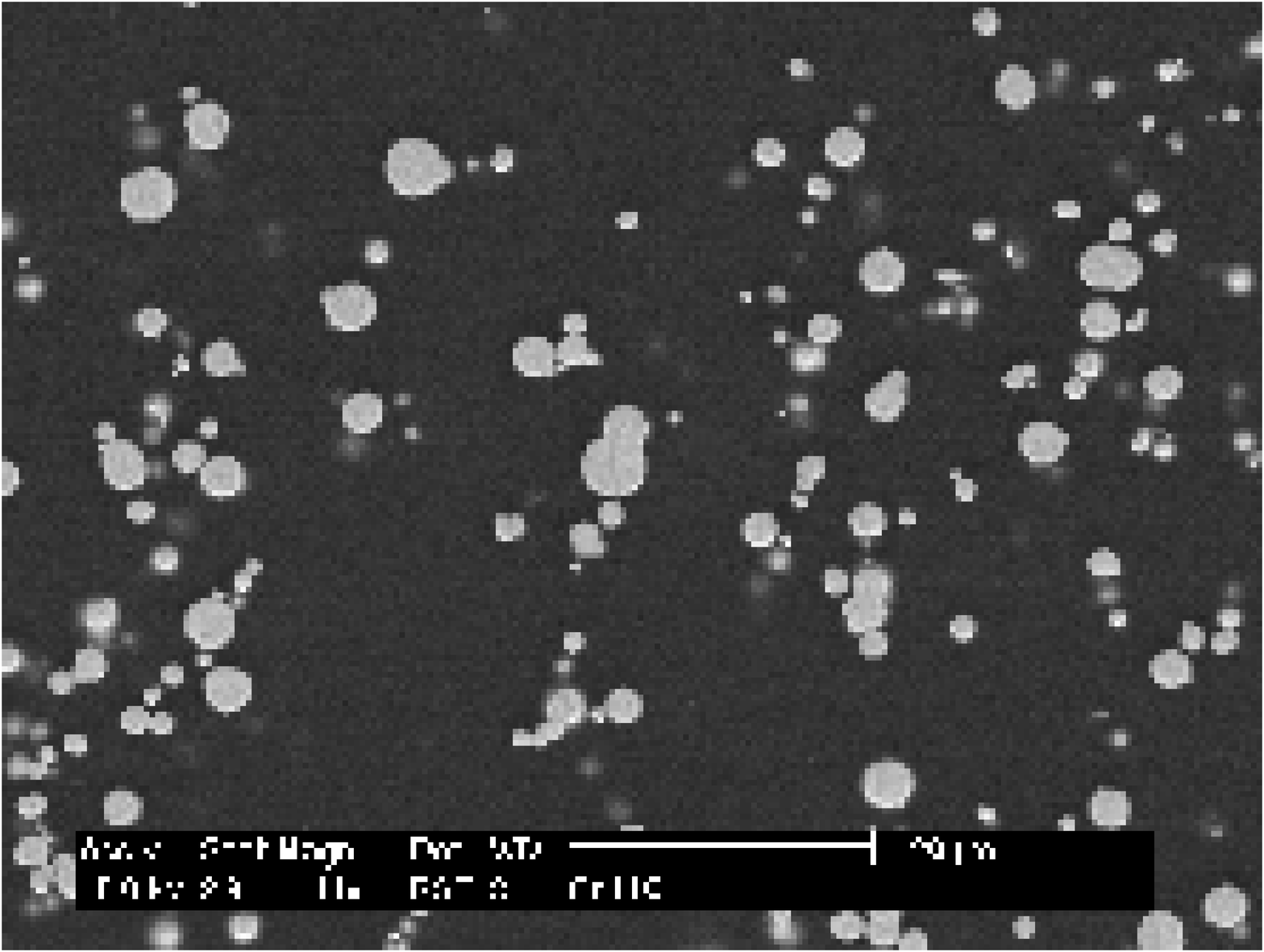,width=9.cm}}
	\caption{Scanning Electron Microscope image of ECCOSORB\texttrademark CR-110 obtained by the authors.}
\label{fig:SEM_CR110}
\end{figure}

\begin{figure}[hp]
	\centerline{\epsfig{file=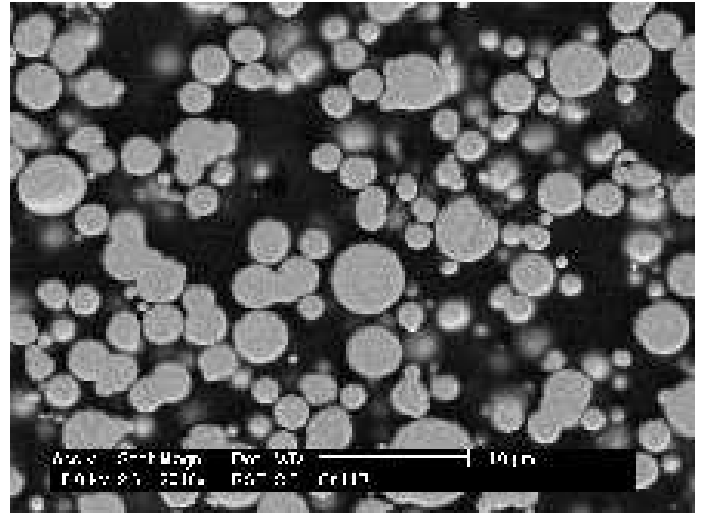,width=9.cm}}
	\caption{Scanning Electron Microscope image of ECCOSORB\texttrademark CR-117. The higher iron powder density is evident.}
\label{fig:SEM_CR117}
\end{figure}

\begin{table*}
\begin{minipage}[t]{\columnwidth}
	\caption{Eccosorb CR-110 and CR-117 mechanical, thermal and outgassing properties. Data refer to 300K, unless where indicated}
		\label{tab:Eccosorbmechanical}
		\centering
		\begin{tabular}{lccc}
			\hline\hline
			{\protect{\bf ECCOSORB\texttrademark}} &&&\\
			& CR - 110 & CR - 117 & \\
			&&& \\
			
			Density $(g/cm^3)$ & 1.6 &  4.1  & Emerson \& Cuming \\
			CTE (1/K) & $3\times 10^{-5}$ &  & this work\\
			Elastic Mod. (MPa)& 3942 &  14336 &  this work\\
			Thermal conductivity (W/(m K)) &0.08 & &\cite{1986ApOpt..25..565H}\\
 		 	Specific heat (J/(Kg K))&  0.6 ${\rm T}^{2.05}$ & & \cite{1984IJIMW...5.1507P}\\
			&&&\\
			Outgassing &&& \\
			TML (\%) & 0.44 & & \cite{ESA_outgassing} \\
			RML (\%) &0.23 &   & \cite{ESA_outgassing}  \\
			CVCM (\%) &0.0  &  & \cite{ESA_outgassing}  \\
		\hline
		\end{tabular}
		\end{minipage}
\end{table*}

Eccosorb and Al samples, bonded with epoxy adhesives, were submitted to lap-shear test (test run by ESA). Four epoxy resins were tested: Armstrong A-12, Loctite Hysol EA9394, Loctite Hysol EA9361 and ECCOSORB CR110 (used as adhesive).
Some samples had also been cycled 7 times at 77K. Results are reported in Table \ref{tab:adhmec}.

\begin{table*}
\begin{minipage}[t]{\columnwidth}
	\caption{Mechanical properties of samples bonded with various epoxy adhesives. Data have been measured by ESA Material division.}
		\label{tab:adhmec}
		\centering
		\renewcommand{\footnoterule}{}  
		\begin{tabular}{lcccccc}
			\hline\hline
			Support & Adhesive type & \multicolumn{2}{c}{Tensile stress (MPa)} & \multicolumn{2}{c}{Tensile strain (\%)} & failure mode\\
			& & cycled & not cycled & cycled & not cycled & \\
			\hline
			CR-117/Al & A-12\footnote{mixing ratio 1:1 in weight cured 6h at 55C} & 15.79 & 33.42  & 14.7 & 18.6 & in substrate\\
			CR-117/CR-110 &  & 38.22 & 40.51 & 19.8 & 18.0& in substrate\\
			Al/Al &  & - & 27.94 & - & 0.92 & in adhesive\\
			\hline
			CR-117/Al & Hy9394\footnote{mixing ratio 100:17 in weight, cured 2h at 55C} & 46.29 & 44.76 & 13.7 & 12.1 & in substrate\\
			CR-117/CR-110 &  & 45.34 & 45.36 & 14.0 & 17.4 & in substrate\\
			Al/Al &  & - & 41.12 & - & 13.6 & in adhesive\\
			\hline
			CR-110/Al & CR-110\footnote{mixing ratio 100:12 in weight, cured 4h at 120C}  & 7.05 & 21.4 & 4.7 & 8.2 & in adhesive\\
			CR-110/CR-110 &  & 35.80 & 32.08 & 14.5 & 10.8 & in substrate\\
			\hline
			CR-110/Al & Hy9361\footnote{mixing ratio 100:1140 in weight, cured 2h at 65C}  & 23.47 & 22.21 & 8.3 & 7.6 & in adhesive\\
			CR-110/CR-110 &  & 31.96 & 34.09 & 11.5 & 11.9 & in adhesive\\
			\hline
		\end{tabular}
\end{minipage}
\end{table*}

Hysol EA9394 was selected for the superior performance.

\end{document}